\documentclass[twocolumn]{aastex631}

\usepackage{latexsym}
\usepackage{amssymb}
\usepackage{gensymb}

\usepackage{color}
\definecolor{lightred}     {rgb}{1.00, 0.35, 0.05}
\definecolor{darkred}      {rgb}{0.70, 0.05, 0.05}
\definecolor{lightblue}    {rgb}{0.05, 0.35, 1.00}
\definecolor{blue}         {rgb}{0.03, 0.25, 0.85}
\definecolor{darkblue}     {rgb}{0.05, 0.05, 0.70}
\definecolor{green}        {rgb}{0.05, 0.65, 0.15}
\definecolor{darkgreen}    {rgb}{0.10, 0.55, 0.15}

\usepackage{natbib}

\def \kms{km$\,\rm s^{-1}$~}
\def \kmsn{km$\,\rm s^{-1}$}
\def \kmspc{km$\,\rm s^{-1}$pc$^{-1}$}
\def \micron{{$\mu {\rm m} $}}

\begin{document}
\title{Impacts of bar-driven shear and shocks on star formation}


\author[0000-0002-5857-5136]{Taehyun Kim}
\affiliation{Department of Astronomy and Atmospheric Sciences, Kyungpook National University, Daegu 702-701, Republic of Korea,  tkim.astro@gmail.com, mgp@knu.ac.kr}

\author{Dimitri A. Gadotti}
\affiliation{Centre for Extragalactic Astronomy, Department of Physics, Durham University, South Road, Durham DH1 3LE, UK}

\author{Miguel Querejeta}
\affiliation{Observatorio Astron{\'o}mico Nacional, C/Alfonso XII 3, Madrid 28014, Spain}

\author{Isabel P{\'e}rez}
\affiliation{Departamento de F{\'i}sica Te{\'o}rica y del Cosmos, Campus de Fuentenueva, Universidad de Granada, 18071 Granada, Spain}
\affiliation{Instituto Carlos I de F{\'i}sica Te{\'i}rica y Computacional, Facultad de Ciencias, 18071 Granada, Spain}

\author{Almudena Zurita}
\affiliation{Departamento de F{\'i}sica Te{\'o}rica y del Cosmos, Campus de Fuentenueva, Universidad de Granada, 18071 Granada, Spain}
\affiliation{Instituto Carlos I de F{\'i}sica Te{\'i}rica y Computacional, Facultad de Ciencias, 18071 Granada, Spain}

\author{Justus Neumann}
\affiliation{Max Planck Institute for Astronomy, K{\"o}nigstuhl 17, 69117, Germany}

\author{Glenn van de Ven}
\affiliation{Department of Astrophysics, University of Vienna, T{\"u}rkenschanzstra{\ss}e 17, 1180 Vienna, Austria}

\author{Jairo M{\'e}ndez-Abreu}
\affiliation{Instituto de Astrof{\'i}sica de Canarias, Calle V{\'i}a L{\'a}ctea s/n, 38205 La Laguna, Tenerife, Spain}
\affiliation{Departamento de Astrof{\'i}sica, Universidad de La Laguna, 38200 La Laguna, Tenerife, Spain}

\author{Adriana de Lorenzo-C{\'a}ceres}
\affiliation{Instituto de Astrof{\'i}sica de Canarias, Calle V{\'i}a L{\'a}ctea s/n, 38205 La Laguna, Tenerife, Spain}
\affiliation{Departamento de Astrof{\'i}sica, Universidad de La Laguna, 38200 La Laguna, Tenerife, Spain}

\author{Patricia S{\'a}nchez-Bl{\'a}zquez}
\affiliation{Departamento de F{\'i}sica de la Tierra y Astrof{\'i}sica, Universidad Complutense de Madrid, E-28040 Madrid, Spain}
\affiliation{Instituto de F{\'i}sica de Part{\'i}culas y del Cosmos (IPARCOS), Universidad Complutense de Madrid, E-28040}

\author{Francesca Fragkoudi}
\affiliation{Institute for Computational Cosmology, Department of Physics, Durham University, South Road, Durham DH1 3LE, UK}

\author{Lucimara P. Martins}
\affiliation{Universidade Cidade de S{\~a}o Paulo/Universidade Cruzeiro do Sul, Rua Galv{\~a}o Bueno 868, S{\~a}o Paulo-SP, 01506-000, Brazil}

\author{Luiz A. Silva-Lima}
\affiliation{Universidade Cidade de S{\~a}o Paulo/Universidade Cruzeiro do Sul, Rua Galv{\~a}o Bueno 868, S{\~a}o Paulo-SP, 01506-000, Brazil}

\author{Woong-Tae Kim}
\affiliation{Department of Physics and Astronomy, Seoul National University, Seoul 08826, Republic of Korea}
\affiliation{SNU Astronomy Research Center, Seoul National University, Seoul 08826, Republic of Korea}

\author{Myeong-gu Park}
\affiliation{Department of Astronomy and Atmospheric Sciences, Kyungpook National University, Daegu 702-701, Republic of Korea,  tkim.astro@gmail.com, mgp@knu.ac.kr}

\begin{abstract}
Bars drive gas inflow. As the gas flows inwards, shocks and shear occur along the bar dust lanes. Such shocks and shear can affect the star formation and change the gas properties. For four barred galaxies, we present H$\alpha$ velocity gradient maps that highlight bar-driven shocks and shear using data from the PHANGS-MUSE and PHANGS-ALMA surveys which allow us to study bar kinematics in unprecedented detail. Velocity gradients are enhanced along the bar dust lanes, where shocks and shear are shown to occur in numerical simulations. Velocity gradient maps also efficiently pick up expanding shells around HII regions. 
We put pseudo slits on the regions where velocity gradients are enhanced and find that H$\alpha$ and CO velocities jump up to $\sim$ 170 \kmsn, even after removing the effects of circular motions due to the galaxy rotation. 
Enhanced velocity gradients either coincide with the peak of CO intensity along the bar dust lanes or are slightly offset from CO intensity peaks, depending on the objects. Using the BPT diagnostic, we identify the source of ionization on each spaxel and find that star formation is inhibited in the high velocity gradient regions of the bar, and the majority of those regions are classified as LINER or composite. This implies that star formation is inhibited where bar-driven shear and shocks are strong.
Our results are consistent with the results from the numerical simulations that show star formation is inhibited in the bar where shear force is strong.
\end{abstract}
\keywords{Barred spiral galaxies(136); Galaxy kinematics (602); Star formation (1569); Galaxy structure (622); Spiral galaxies (1560)}

\section{Introduction} 
Numerical simulations and observations find that once a galactic disk is massive enough and dynamically cold, it is often easy to form bars (e.g., \citealt{hohl_71, ostriker_73, combes_81, athanassoula_02a, kraljic_12, sheth_12, seo_19, fragkoudi_20}).
Bars are frequently found in nearby galaxies, and the fraction of barred galaxies exceeds 65\% including weak bars (e.g., \citealt{devaucouleurs_63, eskridge_00, menendez_delmestre_07, barazza_08, aguerri_09, masters_11, buta_15, diaz-garcia_16,lee_19}).
Bars are non-axisymmetric structures inducing tangential forces and gravitational torques on the gas. As the gas around the bar within the corotation radius loses angular momentum, it flows inward.
Piled-up gas in the central region ignites active star formation (SF, \citealt{hawarden_86, knapen_95, sheth_05}) to form nuclear disks which lead to the formation of inner bars in the nuclear disks and nuclear rings in some galaxies (e.g., \citealt{sormani_18b, seo_19, delorenzo_caceres_19b, gadotti_20, bittner_21, desafreitas_23a, desafreitas_23b}). 
The gas transported to the central 100 pc may feed active galactic nuclei (e.g., \citealt{storchi-bergmann_19, silva_lima_22, garland_23, kolcu_23}), although the connection between bars and AGN is still controversial (e.g., \citealt{ho_97, knapen_00, laine_02, cisternas_13, cheung_15, neumann_19, zee_23}).
Moreover, bars also enrich the diversity of the outer part of galaxies. Bars are frequently accompanied by an inner and outer ring and also induce spiral arms (e.g., \citealt{salo_10, comeron_14, buta_15}). 

In addition to structure and morphology, there are significant differences in the observed kinematics between barred and unbarred disk galaxies. Unbarred disk galaxies show iso-velocity maps that can be well explained by circular motions (e.g., \citealt{bosma_81}), while barred galaxies exhibit characteristic ``S-shaped'' bends along the bar in iso-velocity diagrams (e.g., \citealt{peterson_80, fathi_05, lopez_coba_22}).
Due to the non-axisymmetric structure, both gas and stellar kinematics are affected. Gas in the bar region experiences non-circular, streaming motions which is estimated to be $\sim$ 40$-$100 \kmsn (\citealt{rozas_00, zurita_04}). 
Stars in the bar region are thought to follow x1 orbits which are elongated along the major axis of the bar, and also x2 orbits that are rounder, and elongated perpendicular to the bar (e.g., \citealt{contopoulos_80b}).
\begin{figure}[ht!]
\includegraphics[width=0.5\textwidth]{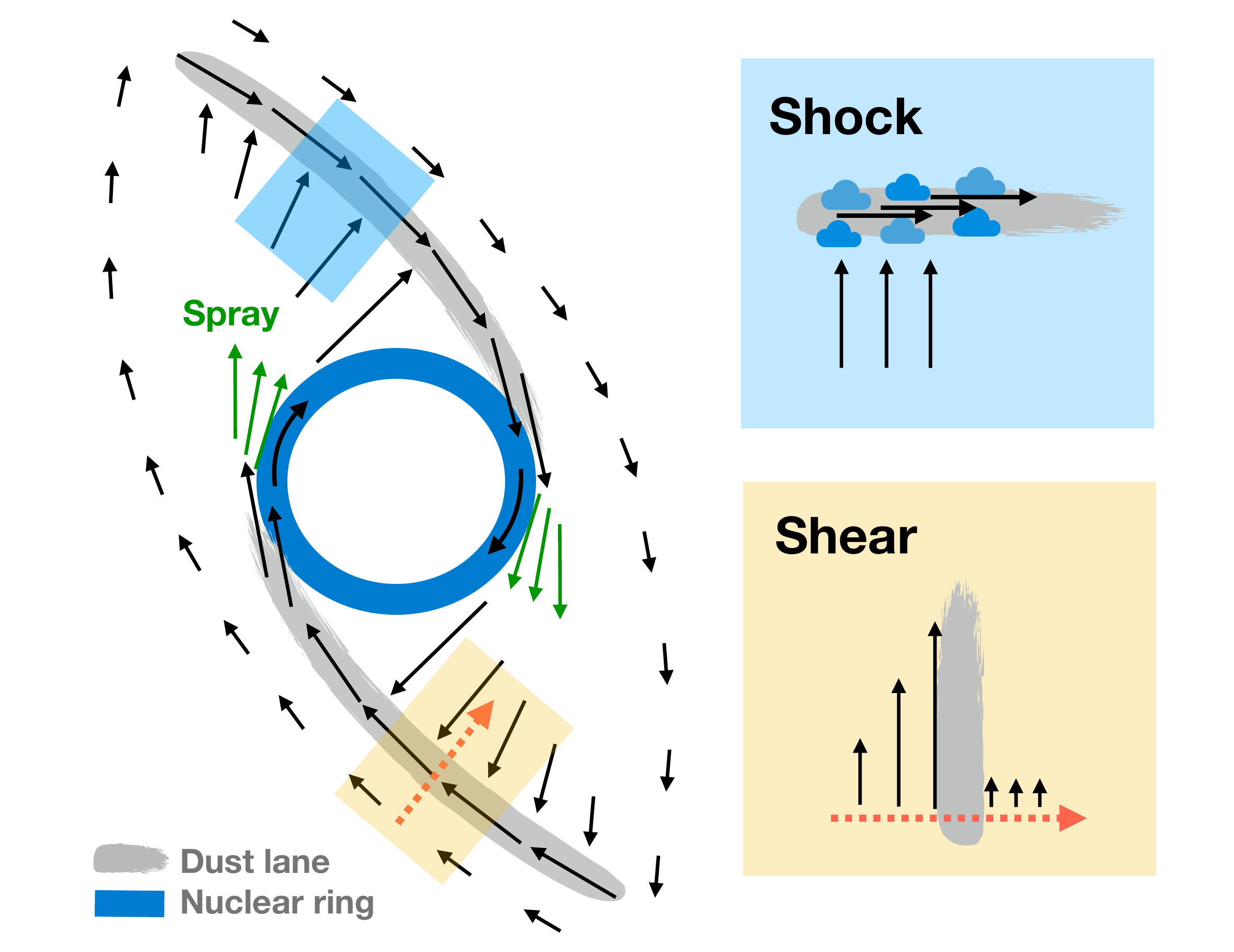}
\caption{Schematic diagram of gas dynamics in the bar region. The magnitude and direction of gas particles are denoted by black arrows. 
Gas particles stream along the bar dust lane (thick gray lines). Some of them flow into the nuclear ring (blue circle), while others are overshoot and sprayed into the other dust lane. 
On the right, the bar-driven shocks and shear are described. 
Gas particles are compressed in the dust lane. As gas enters the dust lane, the gas velocity is changed abruptly in direction and magnitude and thus experiences shocks.
In the lower right panel, only the gas velocity vectors that are parallel to the dust lane are plotted. Shear occurs when there is a velocity gradient across the flow, perpendicular to the dust lane (orange dotted arrow) in this case. }
\label{fig:schematic}
\end{figure}

In Figure \ref{fig:schematic}, we present a schematic diagram that describes the direction and magnitude of gas velocities within the bar region. Gas flows inward along the bar dust lanes which are found at locations of hydrodynamic shocks based in simulations (e.g., \citealt{roberts_79, athanassoula_92b, sormani_15b, fragkoudi_17a}).
If there is a nuclear ring, a part of the gas ($\sim$ 20 -- 30\%, \citealt{regan_97a, hatchfield_21}) enters the nuclear ring while the remaining gas overshoots and sprays toward the other bar dust lane (e.g., \citealt{sormani_19}). On the right panel of Figure \ref{fig:schematic}, shocks and shear are described in the blue and orange boxes, respectively.
When the gas particles are sprayed and encounter the other bar dust lane, the gas particles are suddenly decelerated, and both the direction and the magnitude of the gas velocities are abruptly changed (see e.g., figure 7 and 8 of \citealt{kim_w_12aa}). This mechanism makes dissipative gas particles experience shocks. 
As gas particles flow into the dust lane, they also experience large velocity shear. The shear occurs when there are velocity gradients across the flow, here perpendicular to the dust lane.
To highlight shear motions, only the gas velocity vectors that are parallel to the dust lane are drawn on the right bottom panel of Figure \ref{fig:schematic}. There are velocity gradients along the orange dotted arrow, which lead to velocity shear. More details from simulated barred galaxies can be found in figure 2 of \citet{athanassoula_92b}, figure 10 of \citet{downes_96}, figure 9 of \citet{regan_99}, figure 7(a) of \citet{kim_w_12aa}, and figure 8 and 9 of \citet{sormani_18a}.

Using hydrodynamic simulations, \citet{kim_w_12aa} investigate the formation and properties of various substructures of barred galaxies. They find strong velocity shear across the bar dust lane amounting to $\sim$1--3 \kmspc, which is $\sim$10 times larger than the shear from the Galactic rotation in the solar neighborhood.

Gas flows into the central regions of the galaxy along the bar dust lanes (e.g., \citealt{athanassoula_92b, englmaier_97, kim_w_12aa, fragkoudi_16, sormani_23}), and as the bar-driven shocks and shear occur right at the dust lane, it is one of the most interesting sites to test the influence of shocks and shear on SF.
In this study, we aim to examine whether SF is promoted or suppressed along the bar where strong shear and shocks are expected.

The loci of SF and star formation rate (SFR) vary among barred galaxies (e.g., \citealt{verley_07, neumann_19, diaz-garcia_20, fraser-mckelvie_20a}).
Detailed studies on individual galaxies found that SF in the bar region is low (e.g., \citealt{tubbs_82}), although the gas column densities in the bar dust lanes are high (\citealt{sheth_00}). Star formation efficiency (SFE) is found to be low in the bar region compared to other regions in the disk (\citealt{Momose_10, hirota_14, muraoka_16, yajima_19, maeda_20b,pessa_22, maeda_23}).

However, studies on larger samples give different results and find that SFE in the bar region is not systematically lower compared to that in the other regions (e.g., \citealt{Muraoka_19, diaz-garcia_21,querejeta_21}).
\citet{diaz-garcia_21} report that SFE in strong bars is not systematically inhibited, and the SFEs are roughly constant along bars. Using PHANGS-ALMA, \citet{querejeta_21} define five environments within galaxies (center, bar, spiral arm, interarm region, and disk without strong spirals) and find that the SFR surface density in bars is comparable to that in spiral arms and higher than those in disks and interarm regions. 

Spatial resolution may play a role in resolving the detailed structures in the bar region (e.g., nuclear rings). Large IFS surveys (e.g., CALIFA, \citealt{sanchez_12}, MaNGA, \citealt{bundy_15}) typically have an average spatial resolution of $\sim$ 1--2 kpc, and therefore often cannot resolve nuclear rings, nuclear discs or other smaller structures or star-forming sites. On the other hand, recent surveys such as TIMER (\citealt{gadotti_19}), PHANGS-ALMA (\citealt{leroy_21a, leroy_21b}) and PHANGS-MUSE (\citealt{emsellem_22}) provide more resolved datasets (with spatial resolution $\lesssim$ 100 pc).
However, different results on SFR and SFE do not seem to be particularly affected by different resolutions. The main difference may arise from the definition of the bar ends. Because the SF in bar ends is often high and efficient for some galaxies (e.g., \citealt{martin_97, hirota_14, renaud_15, beuther_17, yajima_19}), including the bar ends (e.g., \citealt{querejeta_21}) or excluding the bar ends (e.g., \citealt{maeda_23}) in analyzing SFE in bar regions would yield different results. The differences may also stem from various samples of different bar evolutionary phases (e.g., \citealt{verley_07, fraser-mckelvie_20a}).

Several mechanisms have been introduced to explain the suppression of SF in the bar region.
First, bar-driven high shear prevents SF (e.g., \citealt{tubbs_82, athanassoula_92b, zurita_04, kim_w_12aa, emsellem_15, renaud_15}). Studies using simulations find that shear may suppress the growth of gravitational instability to form stars.
Second, cloud-cloud collisions have been found to affect SF (e.g., \citealt{stone_70a, scoville_86}). If the relative velocity of two colliding gas clouds is high, i.e., if the gas clouds enter into the dust lane with high velocity, then the SF can be suppressed (e.g., \citealt{tubbs_82, reynaud_98, fujimoto_14a, maeda_21}).
Third, gas within the bar extent may already have been all consumed into stars, and thus gas is depleted and the current SF is low (\citealt{spinoso_17, james_18}).
Fourth, the fraction of dense gas is related to SF and SFE (\citealt{Solomon_92, Gao_04, Muraoka_09a, Usero_15, bigiel_16}), and
vigorous motions near the dust lane make it difficult for dense gas to survive and may only exist as diffuse gas in the bar region between the center and bar ends (e.g., \citealt{sorai_12, muraoka_16, yajima_19}).
Lastly, turbulence (\citealt{elmegreen_04_araa, mckee_07, haywood_16, khoperskov_18, sun_20}) and magnetic fields found in the dust lanes of the bar (\citealt{beck_99, kim_w_12bb}) may prevent molecular clouds from collapsing to form stars (\citealt{krumholz_19}).

Extensive studies on NGC 1530, an archetypal barred galaxy, have been performed on kinematics.
\citet{reynaud_98} argue that the SF in NGC 1530 is inhibited in dust lanes where shocks and shear are expected to be strong. Using Fabry-P{\'e}rot interferometry in H$\alpha$ on NGC 1530, \citet{zurita_04} presents clear observational evidence that shear inhibits massive SF, whereas shocks act to enhance it. 

Another aim of this paper is to examine whether the emission-line ratios are modified due to the bar-driven shocks and shear.
Several studies show that shocks from stellar winds, stellar feedback, supernova explosions, and galactic outflows can change the optical emission-line ratios (e.g., [OIII]/H$\beta$, [NII]/H$\alpha$, [SII]/H$\alpha$ and [OI]/H$\alpha$, \citealt{shull_79, calzetti_04, hong_13, ho_14}). Shocks with velocities greater than 110 \kms can generate sufficient UV radiation for complete ionization of H and He (\citealt{shull_79}). Depending on the shock velocity and contribution of shock, emission-line ratios can change from the star-forming to the low-ionization nuclear emission-line region (LINER) in the BPT diagram (figure 13 of \citealt{ho_14}). Therefore, we examine the emission-line ratio of the bar-driven shocks and shear region in the BPT diagram in order to see whether bar-driven shocks and shear can change the gas properties.

\begin{figure*}[ht!]
\includegraphics[width=\textwidth]{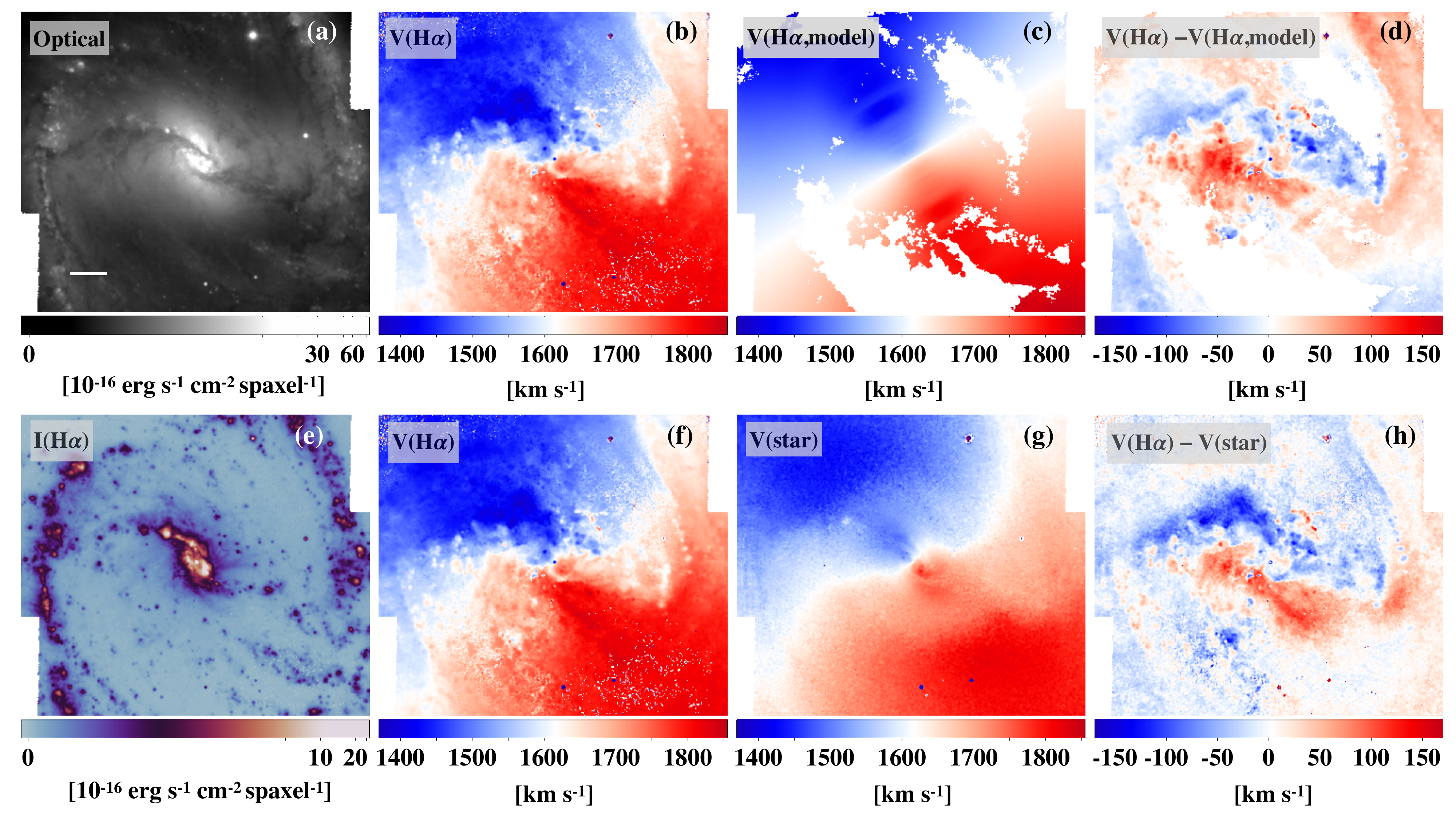}
\caption{NGC 1365 (a): PHANGS-MUSE $5400 \sim 5600 \rm\AA$ summed image. North is up and East to the left. 
(b): H$\alpha$ velocity field, $(V_{\rm H\alpha})$, 
(c): H$\alpha$ velocity model obtained using {\sc 3D--BAROLO}. White regions denote masked regions.
(d): H$\alpha$ residual velocity map, i.e., $(V_{\rm H\alpha} - V_{\rm H\alpha, model})$, obtained by (b)$-$(c),
(e): H$\alpha$ emission-line map,
(f): H$\alpha$ velocity field, $(V_{\rm H\alpha})$, 
(g): stellar velocity field, $V_{\rm star}$, 
(h): velocity difference between H$\alpha$ velocity map and stellar velocity map, $(V_{\rm H\alpha} - V_{\rm star})$ obtained by (f)$-$(g).}
\label{fig:NGC1365_basic}
\end{figure*}

The paper is organized as follows.
We describe the data and the analysis in Section 2.
We explain the methods to estimate the H$\alpha$ velocity gradients to identify bar-driven shock and shear in Section 3.
We quantify velocity jumps across the high velocity gradients in Section 4. 
We investigate whether the bar-induced shocks and shear promote or prevent SF in Section 5.
By examining the BPT map, we classify the ionizing source on each spaxel, and assess the impact of bar-driven shocks and shear on SF in Section 6. 
We conclude and summarize our results in Section 7.

\section{Data} 
\label{sec:data}
We make use of data products from the PHANGS-MUSE survey (\citealt{emsellem_22}). We use the mosaicked datacubes and emission-line fluxes, gas and stellar kinematics, which were convolved and optimized in terms of resolution (``copt''). The PSF of these data have been homogenized spatially across the galaxy, and the seeing of the ``copt'' data of our sample spans from 0.96'' to 1.25''. H$\alpha$ flux and kinematics of the PHANGS-MUSE are obtained using pPXF (\citealt{Cappellari_04}). 
The H$\alpha$ emission-line is fitted treating as an additional Gaussian function, while the stellar continuum is fitted simultaneously. The velocity and velocity dispersion for H$\alpha$ and H$\beta$ are tied together.
Detailed descriptions can be found in \citet{emsellem_22}. The spectral resolution is $\sim$50 \kms at H$\alpha$. CO (2$-$1) emission-line flux and velocity maps are drawn from the PHANGS-ALMA survey (\citealt{leroy_21b}).

Basic properties of our 4 sample galaxies are presented in Table 1.
We present an optical image, maps of H$\alpha$ velocity, galaxy rotation model from H$\alpha$ velocity, difference between H$\alpha$ velocity and galaxy rotation model, H$\alpha$ intensity, stellar velocity, and difference between H$\alpha$ velocity and stellar velocity for NGC 1365 in Figure \ref{fig:NGC1365_basic} as an example.
In order to model the galaxy rotation, we utilize {\sc {3D--BAROLO}} (\citealt{diteodoro_15}) which is a tool for fitting 3D tilted-ring models to emission-line datacubes. 
The position angle, inclination of the galaxies, and center are fixed to the values from Table 2 of \citet{lang_20} in running {\sc3D--BAROLO}. Velocity dispersion and scale height of the disk are fixed to 8 \kms and 0 arcsec, respectively, which are initial values implemented in {\sc {3D--BAROLO}}. An example of rotation model is presented in Figure \ref{fig:NGC1365_basic}(c).
We present velocity differences between the H$\alpha$ velocity and the H$\alpha$ rotation model in Figure \ref{fig:NGC1365_basic}(d), and the velocity differences between H$\alpha$ velocity and stellar velocity maps in Figure \ref{fig:NGC1365_basic}(h).

\begin{deluxetable*}{clccccccccc}
\tablenum{1}
\tablecaption{Basic properties of sample galaxies.}
\tabletypesize{\scriptsize}
\tablehead{
\colhead{Galaxy} & \colhead{Type} & \colhead{Log $M_{\rm star}$} & \colhead{Distance} & \colhead{Log(SFR)} & \colhead{V$_{\rm sys}$} & \colhead{PA$_{\rm disk}$} & \colhead{PA$_{\rm bar}$} & \colhead{Inc} & \colhead{Scale} & \colhead{Copt PSF} 
\\
\colhead{} & \colhead{} & \colhead{[M$_{\odot}$]} & \colhead{[Mpc]} & \colhead{[M$_{\odot}$/yr]} & \colhead{[km$/$s]} & \colhead{[deg]} & \colhead{[deg]} & \colhead{[deg]} & \colhead{[pc/arcsec]} & \colhead{[arcsec]}
}
\decimalcolnumbers
\startdata   
NGC 1365  & $(R')SB(r\underline{s},nr)bc$    & 10.99  & 19.57  & 1.23  & 1613.3  & 201.1  & 88.5 (65$^{\dag}$) & 55.4   & 94.9  & 1.15  \\
NGC 1512  & $(R\underline{L})SB(r,bl,nr)a$   & 10.71  & 18.83  & 0.11  &  871.4  & 261.9  & 43.07              & 42.5   & 91.3  & 1.25  \\ 
NGC 1672  & $(R')SA\underline{B}(rs,nr)b$    & 10.73  & 19.40  & 0.88  & 1319.2  & 134.3  & -84.0              & 42.6   & 94.1  & 0.96   \\ 
NGC 3627  & $SB_{\rm{x}} \rm{(s)b~pec}$    & 10.83  & 11.32  & 0.58  & 717.9   & 173.1  & -22.81             & 57.3   & 54.9  & 1.05   
\enddata
\tablecomments{
(1): Galaxy name.
(2): Mid-Infrared galaxy morphology (\citealt{buta_15}).
(3): Stellar mass (\citealt{leroy_21b}).
(4): Distance (\citealt{anand_21}).
(5): Star formation rate (\citealt{leroy_21b}).
(6): Systemic velocity (\citealt{lang_20}).
(7): Position angle of galaxy from CO kinematics, CCW from the North (\citealt{lang_20}).
(8): Position angle of bar measured using 3.6 $\mu$m images (\citealt{salo_15}).
(9): Inclination (\citealt{lang_20}).
(10): Scale of the PHANGS-MUSE (\citealt{emsellem_22}).
(11): FWHM of the Gaussian PSF of the homogenized, convolved and optimized (copt) PHANG-MUSE mosaic (\citealt{emsellem_22}).  
${\dag}$: the PA of the bar dust lane of NGC 1365. As it is different from the PA of the stellar bar for NGC 1365, we rotate NGC 1365 with the PA of the bar dust lane when we obtain the velocity gradient map. 
}
\end{deluxetable*}

\section{Velocity gradient}
\subsection{Deriving gas velocity gradient maps} \label{sec:deriving_vg}
To explore bar-driven non-circular motions, it is necessary to remove the effect of circular motions.
Non-circular motions can be measured in two ways. The first one is to obtain a galaxy rotation model, and subtract it from the observed velocity field (i.e., the H$\alpha$ velocity field in this study, \citealt{zurita_04}). The second one is to utilize the differences between the gas and stellar velocity field as gas and stars respond to the bar-driven gravitational forces differently. We use both methods to produce velocity gradient maps for NGC 1365 to explain the methodologies and compare them and present results for other galaxies later in this section.
We obtain the velocity gradient map first using the galaxy rotation model with the following steps. 

\begin{figure*}[ht!]
\includegraphics[width=\textwidth]{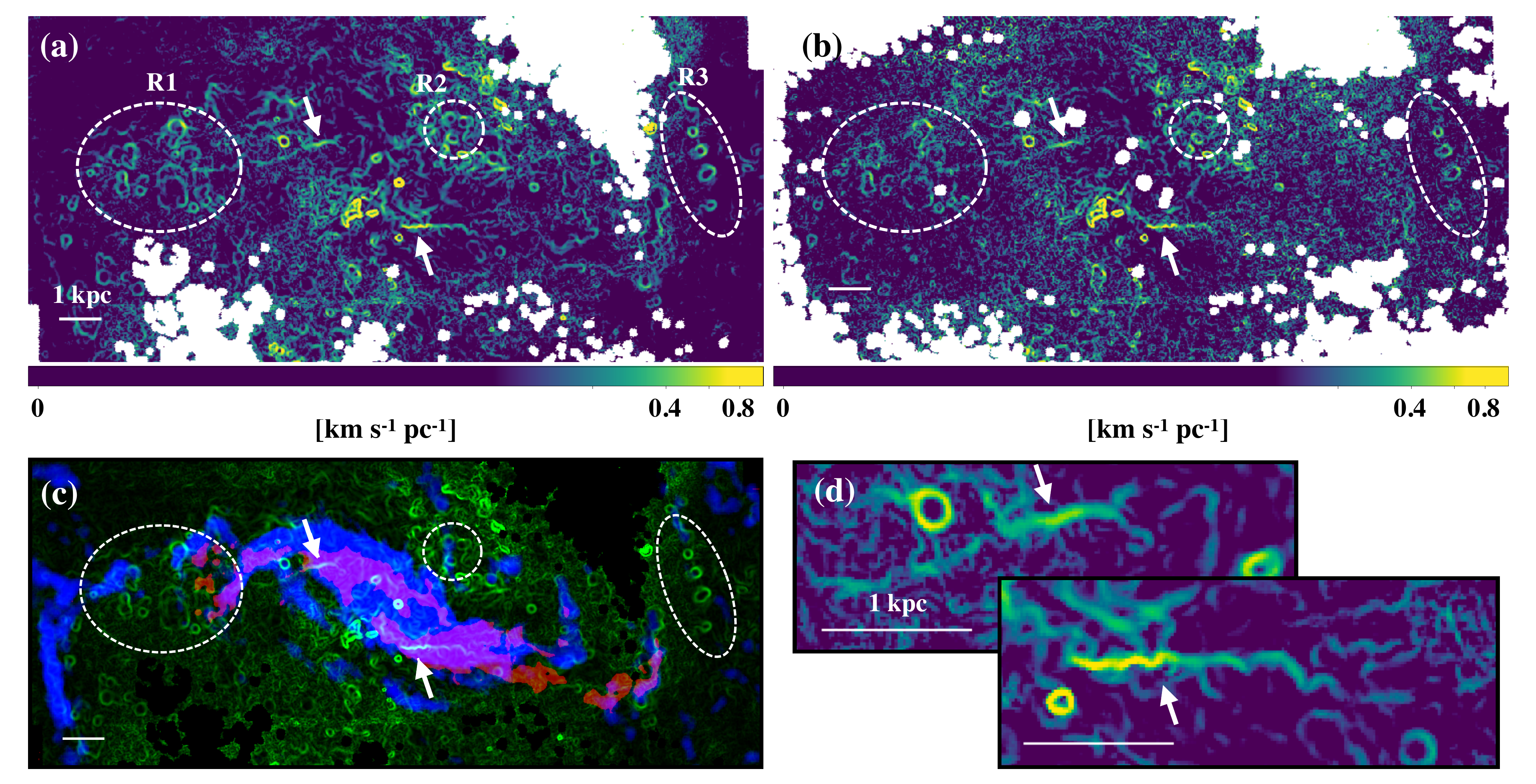}
\includegraphics[width=\textwidth]{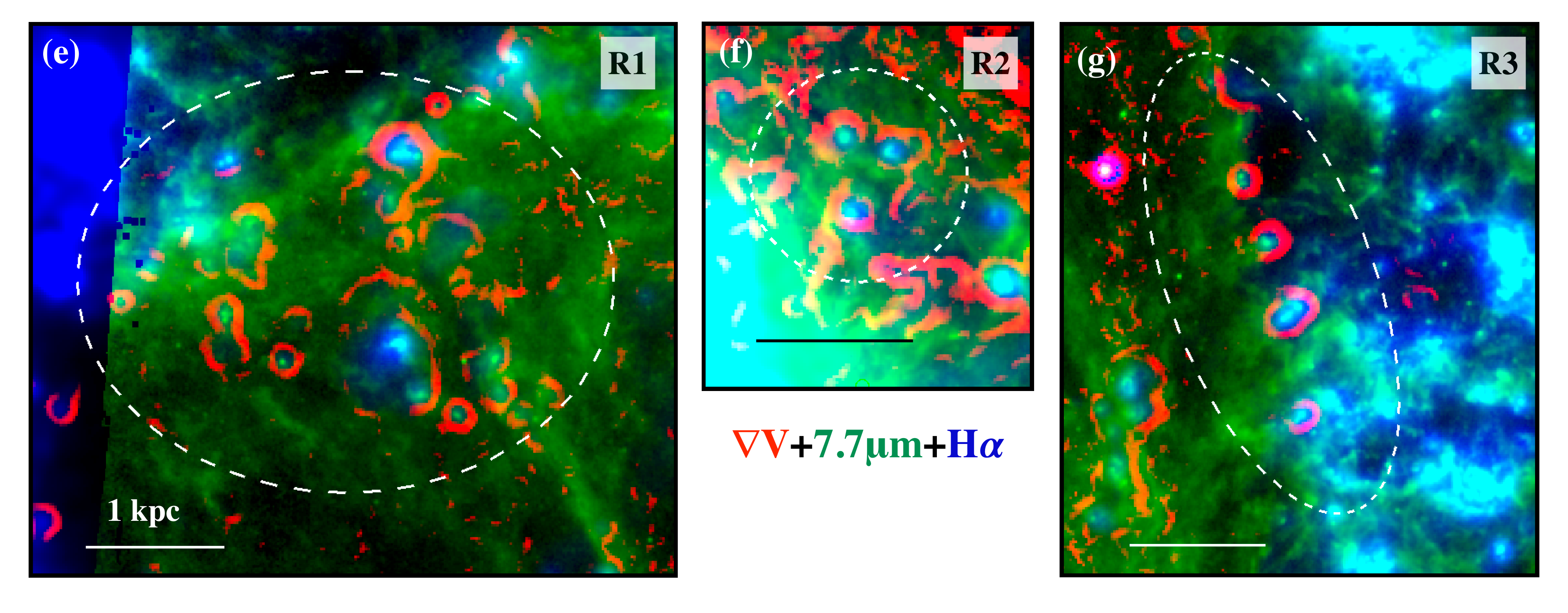}
\caption{Velocity gradient map of NGC 1365. (a): Velocity gradient map calculated using Eq.~\ref{eq:v} on the map of velocity residual of $(V_{\rm H\alpha} - V_{\rm H\alpha, model})$, in unit of [\kmspc]. The white horizontal line in the bottom left spans 1 kpc (10.54$''$). 
(b): Velocity gradient map calculated using the map of $(V_{\rm H\alpha} - V_{\rm star})$.
White arrows denote regions that are prominent in the velocity gradients map which we infer to be driven by bar kinematics. 
White dashed ellipses indicate possible expanding shells around HII regions, supperbubbles, or supernovae.
(c): Velocity gradient map in green, CO $J = 2 - 1$ line emission map from PHANGS-ALMA in blue, and the dust lane mask in red. The dust lane mask is made from the PHANGS-MUSE ancillary data that satisfy the ratio of SDSS g-band flux over r-band flux is less than 0.8. (d): Zoomed-in velocity gradient maps of regions pointed by two arrows.
(e),(f),(g): Zoomed-in RGB images of white dashed ellipses in (a) and (b). Velocity gradients are displayed in red, JWST MIRI F770W images (\citealt{lee_23}) which highlight PAH features at 7.7 \micron ~are shown in green, and PHANGS-MUSE H$\alpha$ emission-line maps are shown in blue. The shells in the figure have a radius in the range of $\sim50$ -- $350$ pc.}
\label{fig:NGC1365_vg_comparison}
\end{figure*}

\begin{enumerate}
\item From the H$\alpha$ velocity map (Figure \ref{fig:NGC1365_basic}b), we subtract the H$\alpha$ galaxy rotation model (Figure \ref{fig:NGC1365_basic}c) to obtain the H$\alpha$ velocity residual map (Figure \ref{fig:NGC1365_basic}d). 

\item We rotate the velocity residual map to align the bar horizontally using the position angle of the bar presented in Table 1, and denote this map as $V_{\rm res}$. 
\item We move along the velocity residual map by 1 pixel (0.2'', 11 $\sim$ 19 pc depending on object) to the left ($V_{\rm res,-1pix}$) and to the right ($V_{\rm res,+1pix}$), and estimate the horizontal velocity gradient as:
\begin{equation}
\nabla V_{x} = \frac{1}{2} \left(| V_{\rm res} - V_{\rm res,-1pix}| + | V_{\rm res} - V_{\rm res,+1pix}| \right)
\label{eq:vx}
\end{equation}
\item We estimate the vertical velocity gradient ($\nabla V_{y}$) similarly by moving 1 pixel up and down, perpendicular to the bar.
\item We add in quadrature both the horizontal ($\nabla V_{x}$) and vertical ($\nabla V_{y}$) velocity gradients to obtain the full velocity gradient.
\begin{equation}
\nabla V = \sqrt{{\nabla V_{x}}^2 + {\nabla V_{y}}^2}
\label{eq:v}
\end{equation}
\end{enumerate}

We plot the velocity gradient map for NGC 1365 in Figure \ref{fig:NGC1365_vg_comparison}(a). 
The unit of the velocity gradient is \kmspc, and it is not corrected for galaxy inclination{\footnote{As we do not intend to compare velocity gradients among different galaxies, velocity gradients are not corrected for inclination. The main conclusion we draw from Section ~\ref{sec:impact_on_SF} and \ref{sec:impact_on_SF_bpt} do not change when correcting for inclination.}}.
\citet{zurita_04} show that velocity gradients perpendicular to the bar trace shear along the bar, while velocity gradients parallel to the bar trace shock along the bar. However, our sample galaxies show rather curved dust lanes. Thus, even velocity gradients perpendicular (parallel) to the bar may not perfectly represent shear (shock). Therefore, we use $\nabla V$ for further analysis. We present $\nabla V_{x}$ and $\nabla V_{y}$ in Appendix \ref{sec:app_vg}.

Velocity gradients can also be estimated using the differences between the H$\rm \alpha$ and observed stellar velocities (Figure \ref{fig:NGC1365_basic}(h)). We calculate the $V_{\rm H\alpha} - V_{\rm star}$ map and follow the same procedure from step 2 to 5 to obtain the velocity gradient map of NGC 1365 (Figure \ref{fig:NGC1365_vg_comparison}(b)).

Comparing Figure \ref{fig:NGC1365_vg_comparison}(a) and (b), we find that the patterns are almost the same while the one from subtracting the galaxy rotation model is more patchy due to low signal-to-noise regions in estimating the galaxy rotation model. 
We found similar results for the other sample galaxies where galaxy rotation models are radially smooth. However, when galaxy rotation models are not smooth, the resulting velocity gradient maps becomes bumpy and thus unsuitable for investigating the velocity gradients. 
Therefore, we opt to use the velocity gradient maps from $V_{\rm H\alpha}-V_{\rm star}$ for further analysis which is unaffected by the galaxy rotation model.
In general, if a galaxy is dynamically hot (e.g., bulge-dominated galaxies), we cannot use stellar velocity fields to trace the galaxy rotation because stars also reside outside the disk plane and are dynamically hotter than gas. However, as our sample of barred galaxies are relatively dynamically cold, Figure \ref{fig:NGC1365_vg_comparison} shows that we can use stellar velocity field to trace the galaxy rotation in order to obtain non-circular motions.

We also have tried to obtain the velocity gradient using CO from the PHANGS-ALMA data set. However, bar regions are not sufficiently detected or are patchy along and adjacent to the bar. Thus it is not suitable to obtain the velocity gradient map over the full bar regions for our sample galaxies, although the spectral resolution is superb. We compare the velocity gradients of H$\alpha$ and CO in Appendix~\ref{sec:vg_co_ha}, and conclude that we can use H$\alpha$ velocity gradients to trace molecular gas velocity gradients for our sample galaxies.

\subsection{Two different types of velocity gradients}

The velocity gradient map tells us where the velocity changes abruptly. As we add vertical and horizontal components in quadrature, it shows velocity change in both directions.
Figure~\ref{fig:NGC1365_vg_comparison} shows that there are various shapes of enhanced velocity gradient regions which can be categorized into two groups.
First, along the bar dust lanes, there are enhanced velocity gradients that are extended more than 1 kpc, as indicated with white arrows in Figure \ref{fig:NGC1365_vg_comparison}a and b. 
These are directly connected to the gas inflow along the bar. Therefore, we claim that these velocity gradients are produced by bar-driven shocks and shear.

Second, there are circular or arc-shaped enhanced velocity gradients that are few hundred parsecs in radius which are not related to the bar but spread throughout the galaxy (see white dashed ellipses in Figure \ref{fig:NGC1365_vg_comparison}). 
The shape of the regions with enhanced velocity gradients suggests they are expanding.
Such shapes of high velocity gradients are especially numerous in NGC 1365. 
We find that H$\alpha$ sources and PAH emissions at 7.7 \micron ~are located at the center of these circles and arcs (see Figure \ref{fig:NGC1365_vg_comparison}).
Thus, we infer that these are due to the expanding shells from HII regions ionized by young stars or stellar associations (\citealt{relano_05, bresolin_20}). Also supernova remnants, superbubbles (\citealt{watkins_23}) or pre-stellar compression shocks may well produce such high velocity gradients.
These velocity gradients are not artificially created by procedures we followed. The velocity difference in these regions can be even seen directly in the H$\alpha$ velocity map and $V_{\rm H\alpha} - V_{\rm star}$ in Figure \ref{fig:NGC1365_basic}(b) and (h). We will study these velocity gradients of expanding regions in more detail in a future paper, while in this work we focus on the first type of high velocity gradients and examine whether the bar-driven shocks and shear inhibit or promote the SF.

\begin{figure*}[h!]
\includegraphics[width=\textwidth]{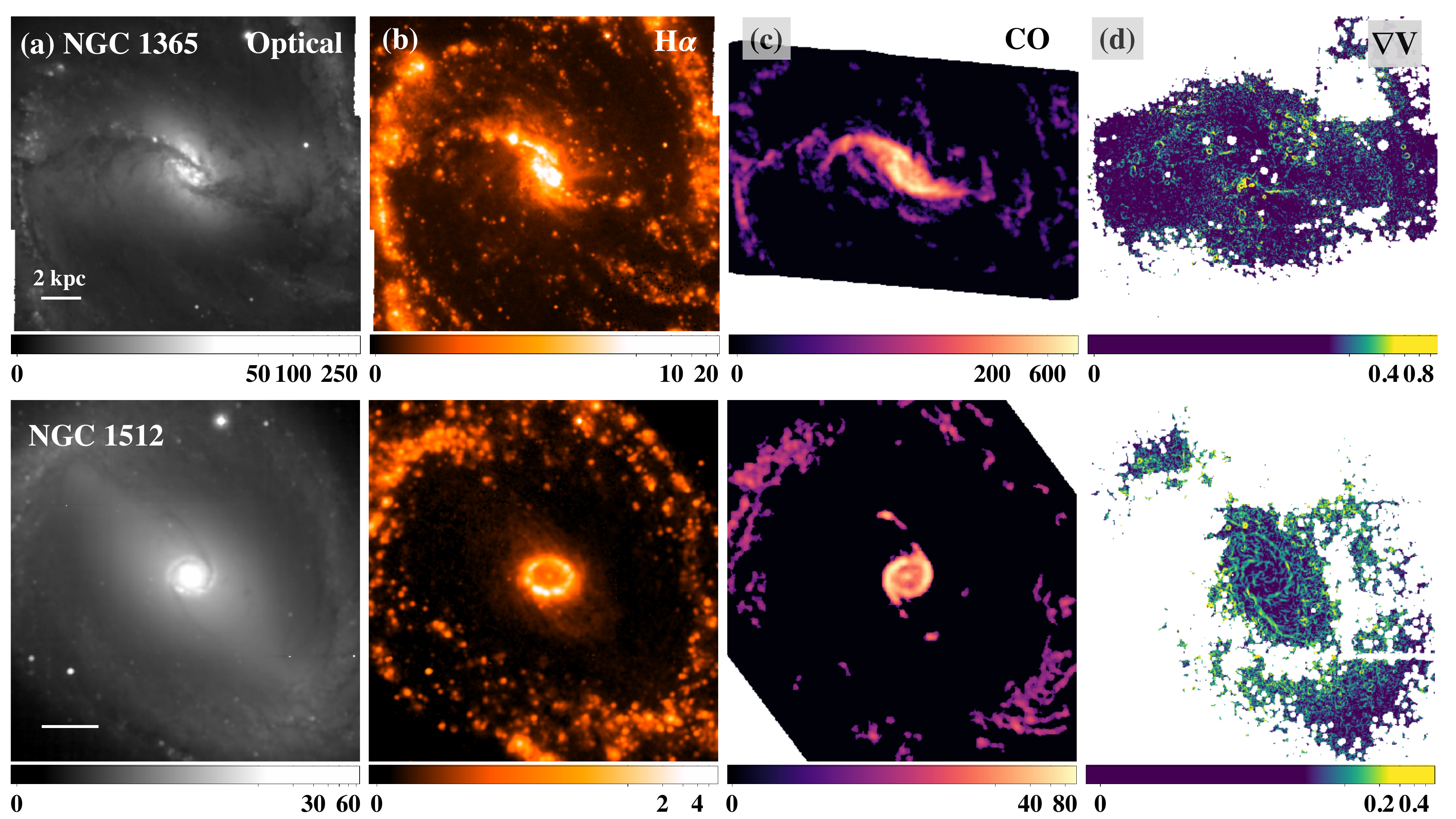} 
\includegraphics[width=\textwidth]{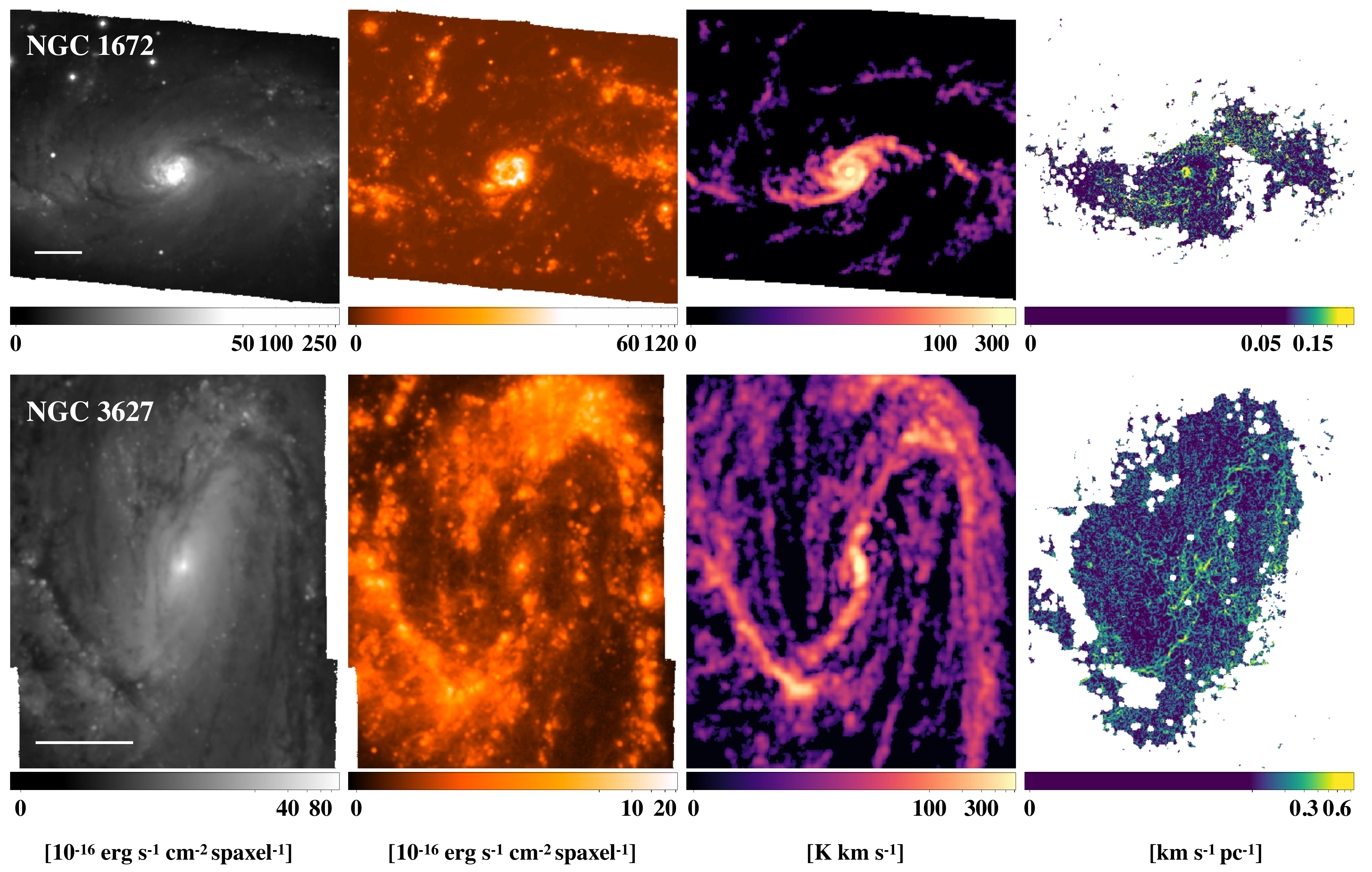} 
\caption{
(a): Optical image from PHANGS-MUSE, $5400 \sim 5600\rm\AA$ summed image. The white line on the bottom left covers 2 kpc for each galaxy. North is up and East to the left.
(b): PHANGS-MUSE H$\alpha$ flux map.
(c): CO $J = 2 - 1$ line emission map from PHANGS-ALMA, and 
(d): velocity gradient map calculated using $(V_{\rm H\alpha} - V_{\rm star})$. Velocity gradients are not corrected for inclinations.}
\label{fig:vg_set4}
\end{figure*}

\subsection{Velocity gradients due to bar-driven gas inflow}
Velocity gradient maps for our sample galaxies are presented in Figure \ref{fig:vg_set4} along with optical, H$\alpha$, and CO images. 
In deriving velocity gradient maps, we applied amplitude-over-noise ratio of H$\alpha$ greater than 3.
We exclude regions where the H$\alpha$ velocity errors ($\delta \rm V_{\rm H\rm \alpha} (x,y)$) are high in the velocity gradient map as follows.
We mask out pixels where $\delta \rm V_{\rm H\rm \alpha} (x,y) > \langle \rm \delta \rm V_{H\alpha} \rangle + C \times\sigma(\delta V_{\rm H\alpha})$, where $\langle \delta V_{\rm H\alpha} \rangle$ is the average of H$\alpha$ velocity errors and $\sigma(\delta \rm V_{\rm H\alpha})$ is standard deviation of $\delta V_{\rm H\alpha}$ for each galaxy. Constant C is set to 0.2 for NGC 1365 and NGC 3627, while C is empirically set to $-0.05$ for NGC 1512 and NGC 1672 to avoid spurious signals around the bar to produce a reliable velocity gradient map. 
The PHANGS-MUSE datasets are Voronoi binned to a target signal-to-noise ratio S/N$=$35 in the continuum range 5300 -- 5500 $\rm\AA$), thus regions with low signals have a large bin size. We found that for some of those bins, the differences between H${\rm \alpha}$ and stellar velocity ($V_{\rm H\alpha} - V_{\rm star}$) are artificially large especially at the edge of those bins due to binning. We empirically find that those pixels can be removed if we remove pixels that have the stellar velocity errors greater than the median of stellar velocity errors of each galaxy.

Enhanced velocity gradients clearly show up along the bar dust lanes and they extend out for a considerable length, almost covering the whole bar except for NGC 1365. The velocity gradients of bar region in NGC 1365 range between 0.1 -- 1.0 \kmspc. This is consistent with the study of \citet{zurita_04}, who find velocity gradients up to 0.35 \kmspc for NGC 1530 and also consistent with the study of \citet{kim_w_12aa}, who find 1 -- 3 \kmspc of shear across the bar dust lane from numerical simulations.
Our results should be considered as a lower limit of the velocity gradient because of the smoothing effect due to the spatial resolution (0.2$"/$pixel) and spectral resolution (2.6 \AA, corresponds to 50 \kmsn at H$\alpha$). 
Detailed notes on each object are available in the Appendix~\ref{sec:append_individuals}.

\subsection{Comparison to the CO Map}
Figure 4 shows that the velocity gradients are pronounced along the bars in all samples of galaxies. 
But exactly where are the velocity gradients enhanced? 
Previous studies have found that the velocity gradients coincide with the dust lane, i.e., on the leading side of the bar (see, e.g., \citealt{zurita_04} and \citealt{feng_22} for studies on NGC 1530 and M31, respectively).

\begin{figure*}[htb!]
\includegraphics[width=\textwidth]{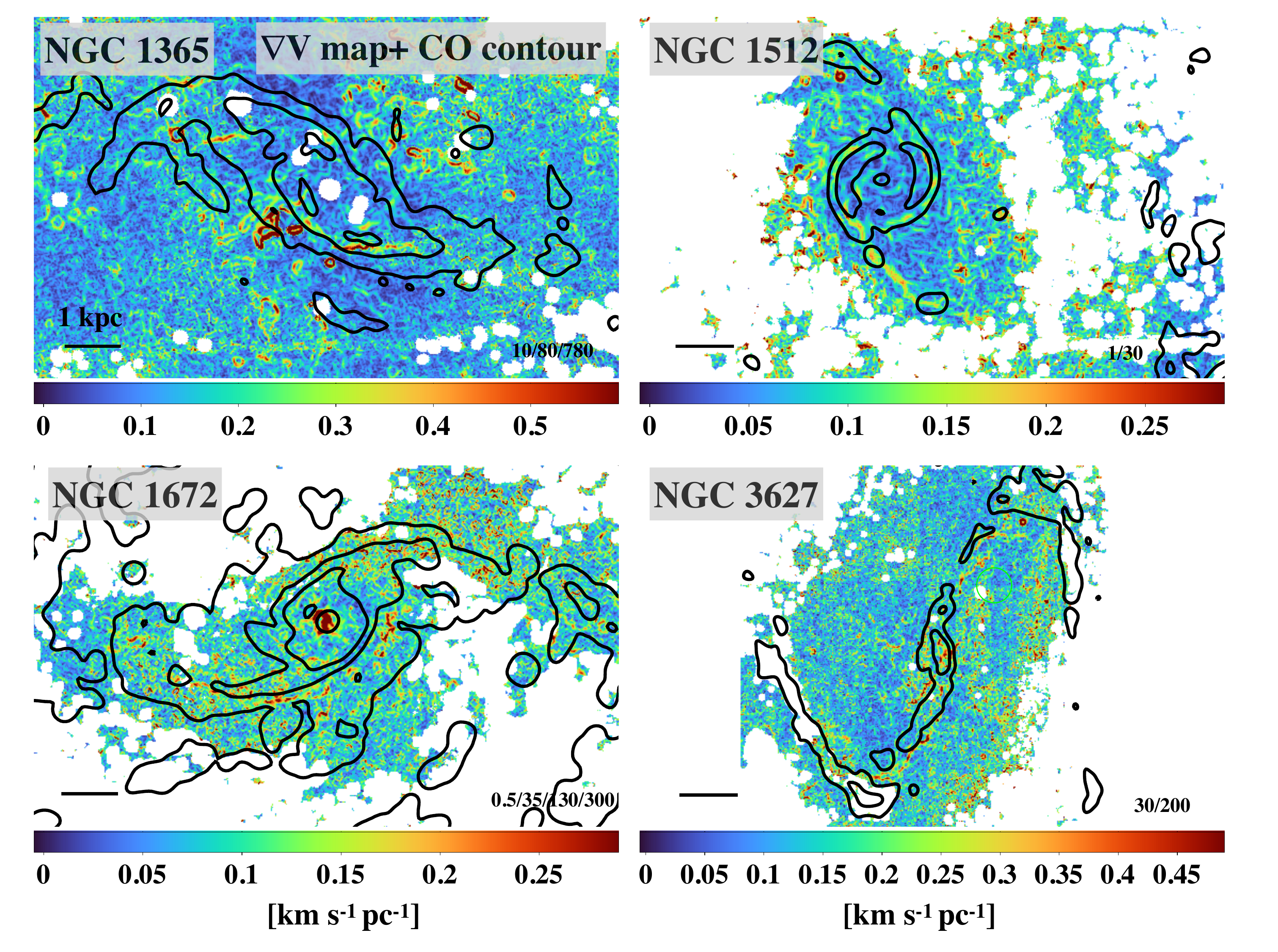} 
\caption{Velocity gradient map superimposed with contours which correspond to the emission-line flux of CO (2$-$1) from the PHANGS$-$ALMA survey. Numbers in the bottom right of each panel correspond to the CO contour levels in [K \kmsn]. The black line on the bottom left covers 1 kpc for each object.}
\label{fig:VG_CO_contour_set4}
\end{figure*}

We plot CO emission-line contours on top of velocity gradient map in Figure \ref{fig:VG_CO_contour_set4} for comparison.
We find that velocity gradients are enhanced either along the bar dust lane inferred from the CO distribution or slightly offset to the edge of the bar dust lane.
In NGC 1365, the southern part of the enhanced velocity gradients are located right next to the brightest contour of the CO emission along the bar dust lane.
We find that in NGC 1512, enhanced velocity gradients coincide with the CO emission detected along the bar dust lane which can be seen in the optical image.
However, in NGC 1672 and NGC 3627, velocity gradients are slightly offset towards to the peak of the CO emission. Velocity gradients are enhanced on the leading side of the bar in NGC 1672 and in the southern part of the NGC 3627. Interestingly, in the northern part of the bar in NGC 3627, it is the opposite. Details on NGC 3627 are in the Appendix \ref{sec:append_individuals}.
In summary, the locations of enhanced velocity gradient regions either coincide with the peak of the CO emissions or are slightly offset to them, depending on the objects.

\begin{figure*}[hbt!]
\includegraphics[width=\textwidth]{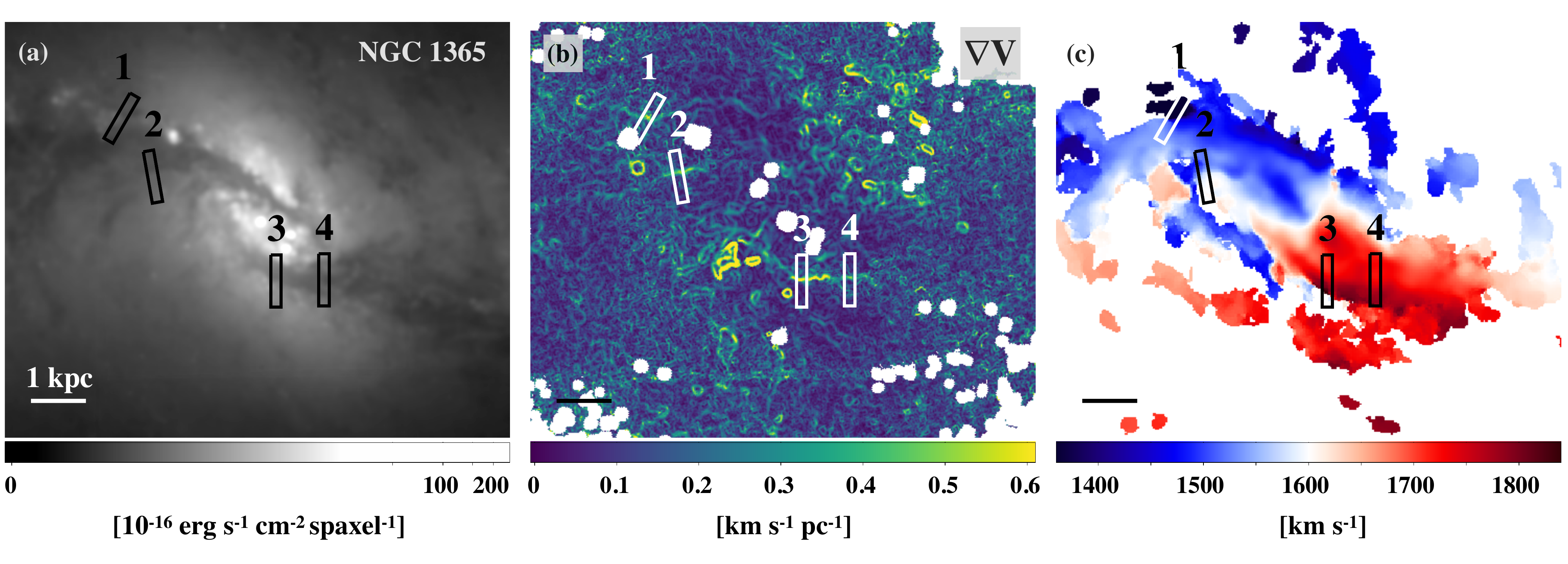}
\includegraphics[width=\textwidth]{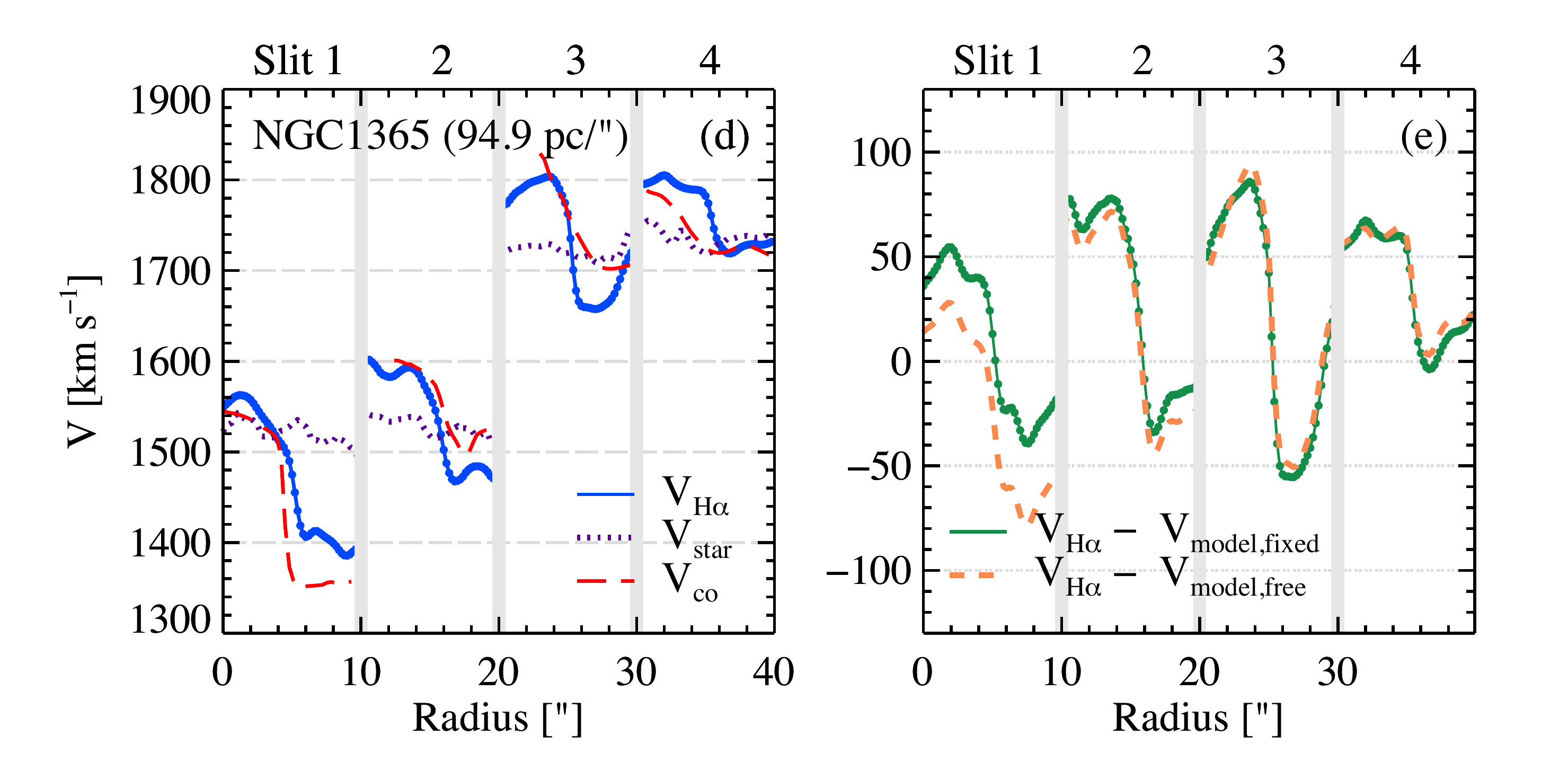} 
\caption{Velocity jumps in high velocity gradient regions of NGC 1365. Four pseudo slits (0.6" $\times$ 10") are put perpendicular to the high velocity gradient segments as in the upper panels of the figure. (a): Optical image of NGC 1365, (b): velocity gradients, (c): CO (2$-$1) velocity, (d) velocity jumps from south to north (upwards) along the slit of H$\alpha$ (blue solid line), CO (red dashed line, if available), and stellar velocity (purple dotted line), (e) velocity jumps after removal of the galaxy rotation model. Two cases of galaxy rotation models are obtained with {\sc 3D--BAROLO}, where position angles and inclinations are fixed (green solid line) and set to free (orange dashed line) for each ring.
Velocity jumps for other sample galaxies are presented in Appendix~\ref{sec:app_veljump}.}
\label{fig:vel_jump}
\end{figure*}

\section{Velocity Jumps}
\label{sec:velocityjump}
We quantify how much velocity jump occurs in the regions where strong bar-driven shocks and shear are expected for our sample galaxies. In order to estimate velocity jumps, we place pseudo slits perpendicular to the regions of high velocity gradient in the bar region.
Figure \ref{fig:vel_jump} shows the velocity changes along the slits in selected regions for NGC 1365.
We find that the velocity jump exceeds 100 \kms within 2''(190 pc) in both H$\alpha$ and CO in the slit 1 and 3. Within 5''(475 pc), the velocity of CO changes up to 200 \kmsn, and that of H$\alpha$ changes up to 150 \kms in slit 1. 
The inclination of NGC 1365 is estimated to be 55.4$^{\circ}$ (\citealt{lang_20}). Thus, the velocity jump would become even higher (${\times}$1.2 for NGC 1365) if we correct for inclination.
We note that these velocity jumps are measured directly from the velocity map, and therefore are not model-dependent. 

While the gaseous velocity jumps are prominent, the stellar velocity jumps are not as pronounced in these regions.
The degree of velocity jumps may be different between gas and stars because they respond differently to the gravitational forces of the bar. 
Gas is dynamically cold and thus highly responsive to the changes in the gravitational potentials it feels as it orbits in the disk.
In addition, gas is collisional and thus easily influenced by shear and shocks, and therefore loses angular momentum more easily, which leads to stronger non-circular motions. However, stars are generally not affected by shocks and stars making up bars are in more stable orbits which lead to milder velocity jumps. 

Components of galactic rotation are included in the velocity map and thus in these velocity jumps. Therefore, to exclude the effects of circular motions from galaxy rotation, we subtract the galaxy rotation model from H$\alpha$ velocity map. Then we assess the velocity jump in the bottom right panel of Figure \ref{fig:vel_jump}. In fitting galaxy kinematics using {\sc 3D--BAROLO}, position angles and inclinations can be fixed or freely modeled ring by ring. We present both results. 
Even after the circular motions are removed, we still find the velocity changes reaching 140 \kms (170 \kms after deprojecion) within 2''(190 pc) in slit 3 and 90 \kms (110 \kms after deprojection) within 4''(380 pc) in slit 1 for NGC 1365.  
Velocity jumps and the position of pseudo slits of other sample galaxies are present in Appendix~\ref{sec:app_veljump}. 
Velocity jumps of other sample galaxies after removal of the galaxy rotation span from 50 to 80 \kms (75 to 95 \kms after deprojection) within 2'' and H$\alpha$ velocity jumps are more pronounced compared to stellar velocity jumps as in NGC 1365. 

Velocity jumps reaching $\sim$100$-$150 \kms have been found using H$\alpha$, [NII], [SII], HI and CO (\citealt{pence_84, reynaud_98, laine_99, mundell_99, weiner_01a, zurita_04}), and even up to 200 \kms (\citealt{zanmar_sanchez_08, feng_22}). Thus, our results are comparable to previous observational studies on the other objects.
With idealized hydrodynamic simulations, \citet{kim_w_12aa} show that the velocity along the slit direction changes abruptly by up to 100 \kms within 100 pc while the velocity perpendicular to the slit direction jumps up to 200 \kms within 100 pc in their Figure 8. It is not straightforward to directly compare the values of velocity jump from simulations with our observational data analysis, because we only measure the line of sight velocity. Nevertheless, overall results are still comparable. 
If we know the pattern speed of the bar and the inclination of the galaxy, 
we can derive gas velocity vector (e.g., \citealt{sormani_23}) by assuming that the gas velocity vector in the bar lanes is parallel to the lane, but this is beyond the scope of this paper. 

\section{Impact of shear and shock on star formation}
\label{sec:impact_on_SF}

\begin{figure*}
\includegraphics[width=\textwidth]{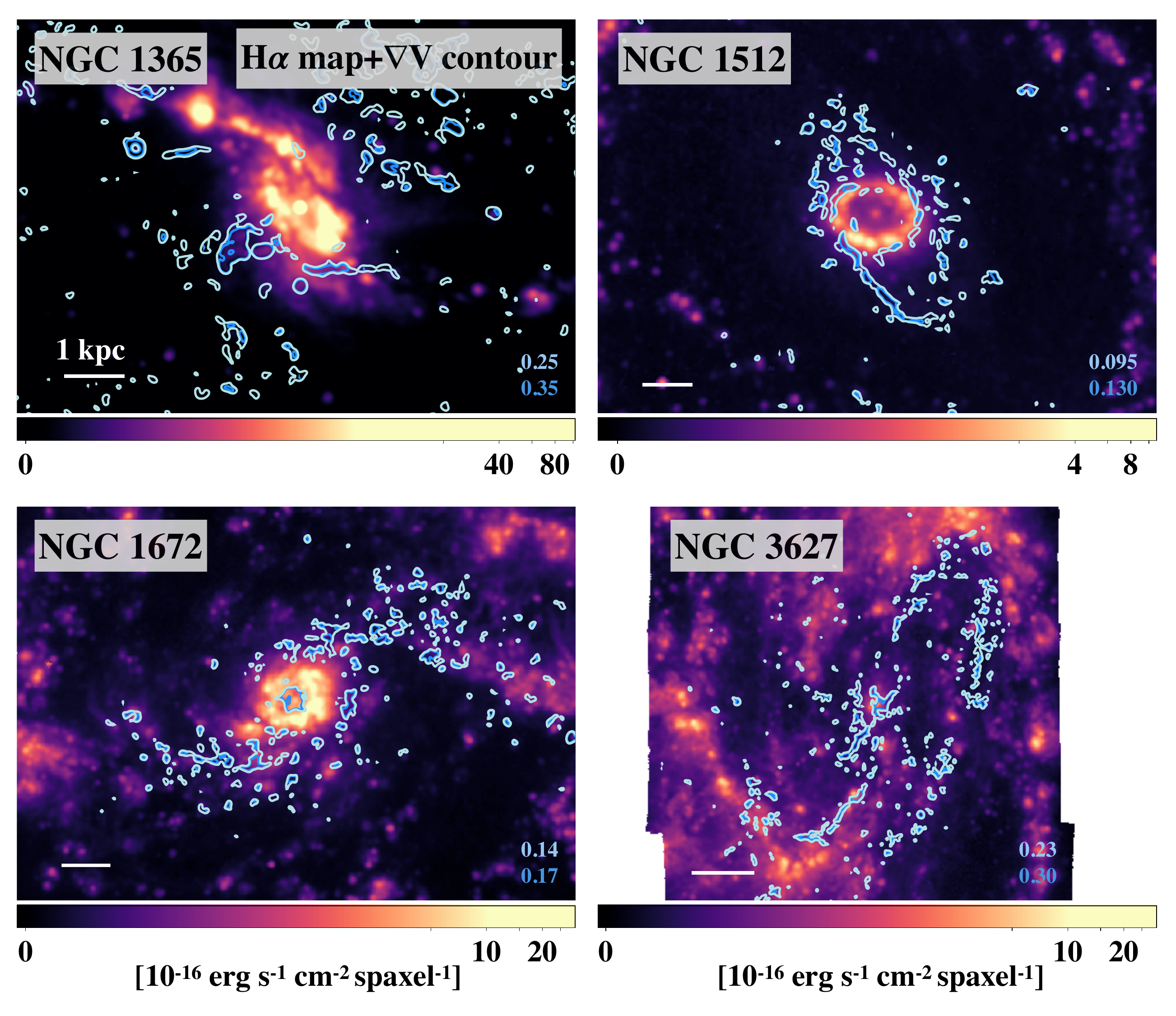}  
\caption{Extinction-corrected H$\alpha$ emission-line maps overlaid with the velocity gradient contours. 
The white line on the bottom of each panel spans 1 kpc. The two numbers on the bottom right in each panel correspond to the velocity gradient contour levels in [km s$^{-1}$ pc$^{-1}$]. Darker blue shades represent higher velocity gradients.}
\label{fig:VG_Ha_contour_set4}
\end{figure*}

\begin{figure*}   
\includegraphics[width=\textwidth]{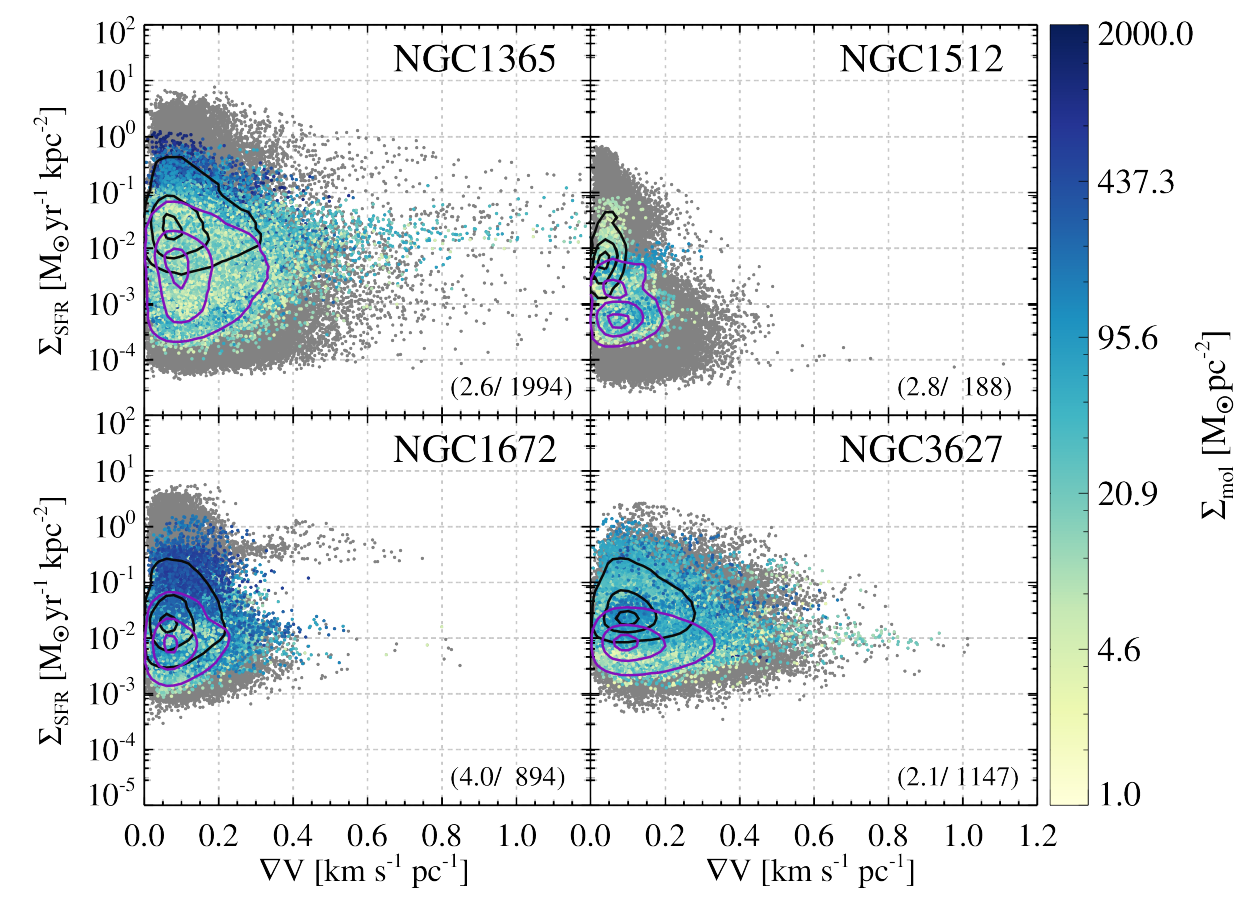}  
\caption{Star formation rate surface density and velocity gradients. Each point corresponds to a spaxel. Bar regions defined by \citet{querejeta_21} (excluding nuclear region and bar ends) are color-coded according to molecular surface density. Spaxels outside the bar regions, and no PHANGS-ALMA CO(2$-$1) detection are plotted in grey. We show the minimum and maximum of the molecular gas surface densities of bar regions in parenthesis for each galaxy. Black contours contain 10$\%$, 50$\%$, and 90$\%$ of star-forming spaxels in the bar region where bar ends and nuclear regions are excluded. Purple contours include 10$\%$, 50$\%$, and 90$\%$ of LINER, Seyfert and composite spaxels in the bar region classified by the BPT diagnostics which are detailed in Section~\ref{sec:impact_on_SF_bpt}.} 
\label{fig:sfrsd_vg}  
\end{figure*}

In this section, we examine the impact of bar-driven shear and shock on star formation which can be traced approximately by velocity gradients and H$\alpha$ emission, respectively. With this aim, we compare the velocity gradients with H$\alpha$ emission-line maps spatially.
Figure \ref{fig:VG_Ha_contour_set4} shows an extinction-corrected H$\alpha$ emission-line map superimposed with the velocity gradient contours. 
To correct for extinction, we use the flux ratio of two emission-lines, H$\alpha$/H$\beta$ (Balmer decrement) (e.g., \citealt{berman_36, groves_12}). 
We assume the value of the intrinsic H$\alpha$/H$\beta$ to be 2.86, corresponding to a temperature $T =10^4$ K and an electron density $n_{e} = 10^2 \rm cm^{-3}$ for Case B recombination (\citealt{baker_38}).
Overall, bright H$\alpha$ sources are rarely found in high velocity gradient regions along the bar.
In particular, SF in NGC 1512 is concentrated in the nuclear ring and the inner ring that encircles the bar, and bright H$\alpha$ sources are rare inside of the ring that encompasses the bar. 
In NGC 3627, there are many star forming regions along the spiral and inter-arm regions. Interestingly, high velocity gradients are avoiding bright star forming regions. 
Overall, bright SF knots are not found on the enhanced velocity gradient regions.

Then, we examine whether there is an anti-correlation between star forming sites and regions of strong shear and shock in a quantitative approach.
We plot the SFR surface density and velocity gradients in Figure \ref{fig:sfrsd_vg}. Each point corresponds to a spaxel. To focus on the shear and shock regions that are directly connected to the gas inflow along the bar, we select the bar regions outside the nuclear region and inside the bar ends, and plot the selected regions in color according to each spaxel's molecular gas surface density in Figure \ref{fig:sfrsd_vg}. Nuclear regions, bar ends, outside the bar regions, and regions with no PHANGS-ALMA CO (2$-$1) detection are plotted in grey. We use the environmental masks from \citet{querejeta_21} to define the bar, bar ends, and nuclear regions.

The SFR is calculated based on H$\alpha$ luminosity. 
H$\alpha$ traces recent SF as 90\% of the H$\alpha$ emission comes from stellar populations younger than 10 Myr (\citealt{kennicutt_12}).
Although the SFR can be better estimated with hybrid estimators in combination with H$\alpha$, MIR and/or FUV data, 
it has been shown that once stellar continuum absorption and dust attenuation are properly corrected, 
there is a good correlation between the SFR derived from H$\alpha$ and the SFR derived from H$\alpha$, UV and MIR data (\citealt{catalan_torrecilla_15}, see also \citealt{belfiore_23}). Furthermore, to conserve the resolution of the H$\alpha$ velocity gradients, we use the SFR derived from H$\alpha$ luminosity in this study.

We estimate the SFR using the following relation (\citealt{calzetti_07, kennicutt_12})
\begin{equation}
\rm{SFR [M_{\odot} yr^{-1}] = 5.3 \times 10^{42} L(\rm H\alpha)_{corr} [erg\,s^{-1}]},
\end{equation}
where $\rm L(\rm H\alpha)_{corr}$ is the extinction-corrected $\rm H\alpha$ luminosity. The molecular mass surface density can be estimated using CO(2$-$1) as
\begin{equation}
\rm{\Sigma_{mol} = \alpha^{1-0}_{CO} \: R^{-1}_{21} \: I_{CO (2-1)} cos \, \it{i} },
\end{equation}
where $\alpha^{1-0}_{\rm CO}$ is the CO(1$-$0) conversion factor, $R^{-1}_{21}$ the CO(2$-$1)$-$to$-$CO(1$-$0) line ratio, and \textit{i} the inclination of a galaxy. We use $\alpha^{1-0}_{\rm CO}$ of the standard Milky Way (4.35 [$\rm M_{\odot} \rm pc^{-2} ($K\,\kms$)^{-1}]$, \citealt{bolatto_13a}), and we use $R^{-1}_{21}=0.65$ (\citealt{leroy_13b, denbrok_21}).

If shear and shocks prevent SF, there should be an anti-correlation between the SFR surface density and the velocity gradients, i.e., the SFR surface density should decrease with the velocity gradients.
However, Figure \ref{fig:sfrsd_vg} shows that while the upper envelope of SFR surface density - velocity gradient decreases, data points fill the area below the upper envelope.
In low velocity gradient regions, there are regions of diverse values of SFR surface density. But as the velocity gradient increases, the upper envelope of SFR surface density decreases. 
If we combine it with molecular gas density, this distribution becomes clear to understand. 
In low velocity gradients, i.e., low shear and shock, regions with high molecular gas density regions show active SF. In the high shear/shock region, active SF is rare. Exceptions are the nuclear regions of NGC 1365 and NGC 1672 where they protrude horizontally in the SFR surface density $-$ velocity gradients diagram. 
Our results imply that SF is hindered in bar-driven shear and shocks.
This is consistent with previous studies (\citealt{reynaud_98, zurita_04}), who conclude that steep velocity gradients prevent molecular gas clouds from condensing to form stars. 

Our results are in line with the study of \citet{emsellem_15}, who find a notable correlation between the regions of newly formed stars and the location of high density gas clumps with low shear, but not with high shear using hydrodynamical simulations of a Milky Way-like galaxy. 
In the next section, we re-visit the impact of bar-driven shear and shock on SF using the BPT classification.

\section{Changes to emission-line ratios due to bar-driven shear and shocks: BPT diagrams}
\label{sec:impact_on_SF_bpt}

\subsection{BPT map}
We examine whether shear and shocks driven by bars lead to any change in the ionized gas properties by inspecting emission-line ratio diagnostic diagrams.
 
\begin{figure}[thb!]
\includegraphics[width=8cm]{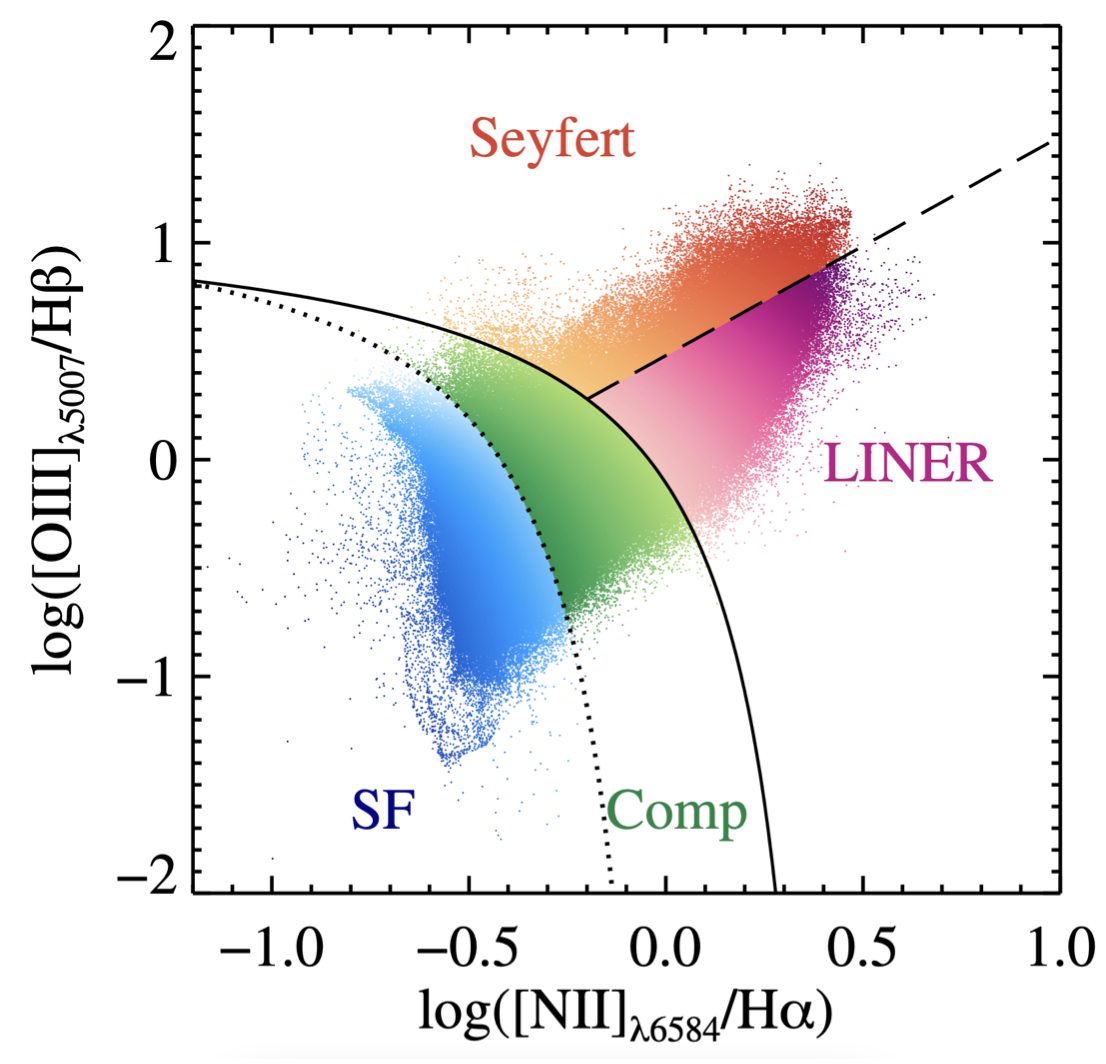} 
\caption{BPT diagram of NGC 1365. Solid line indicates the theoretical maximum starburst line (\citealt{kewley_01}), dotted line marks the limit of pure star formation (\citealt{kauffmann_03c}), and dashed line divides Seyfert and LINER (\citealt{cidfernandes_11}).
Blue points denote emissions from SF regions, green points represent emissions from composite, both from SF and AGN, red to orange points stand for emissions from Seyfert, and magenta points correspond to emissions from LINER. 
Points become darker as they are farther away from the theoretical maximum starburst line (solid line).}
\label{fig:bpt_diagram_1365}
\end{figure}

\begin{figure*}[htb]
\centering
\includegraphics[width=0.8\textwidth]{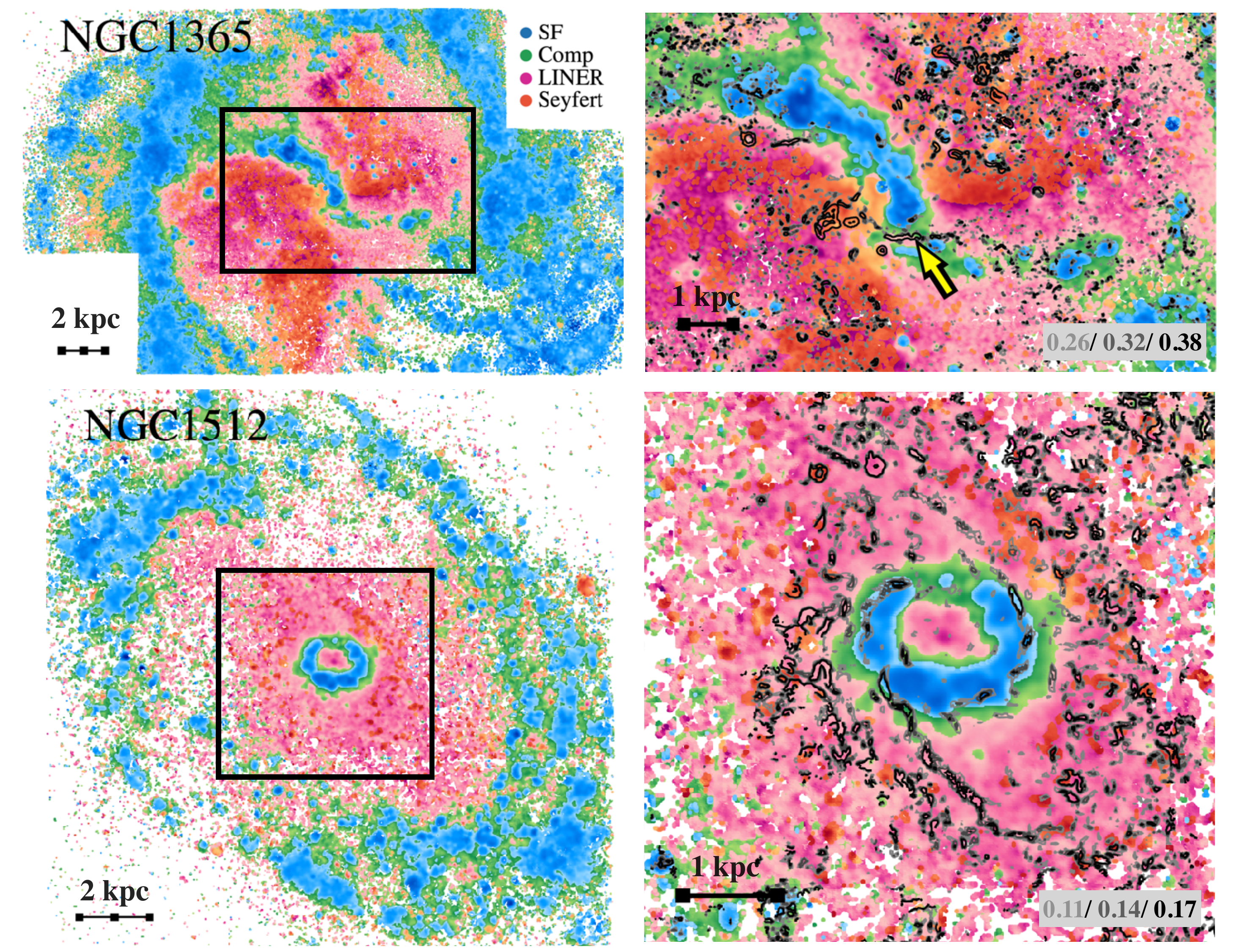} 
\includegraphics[width=0.8\textwidth]{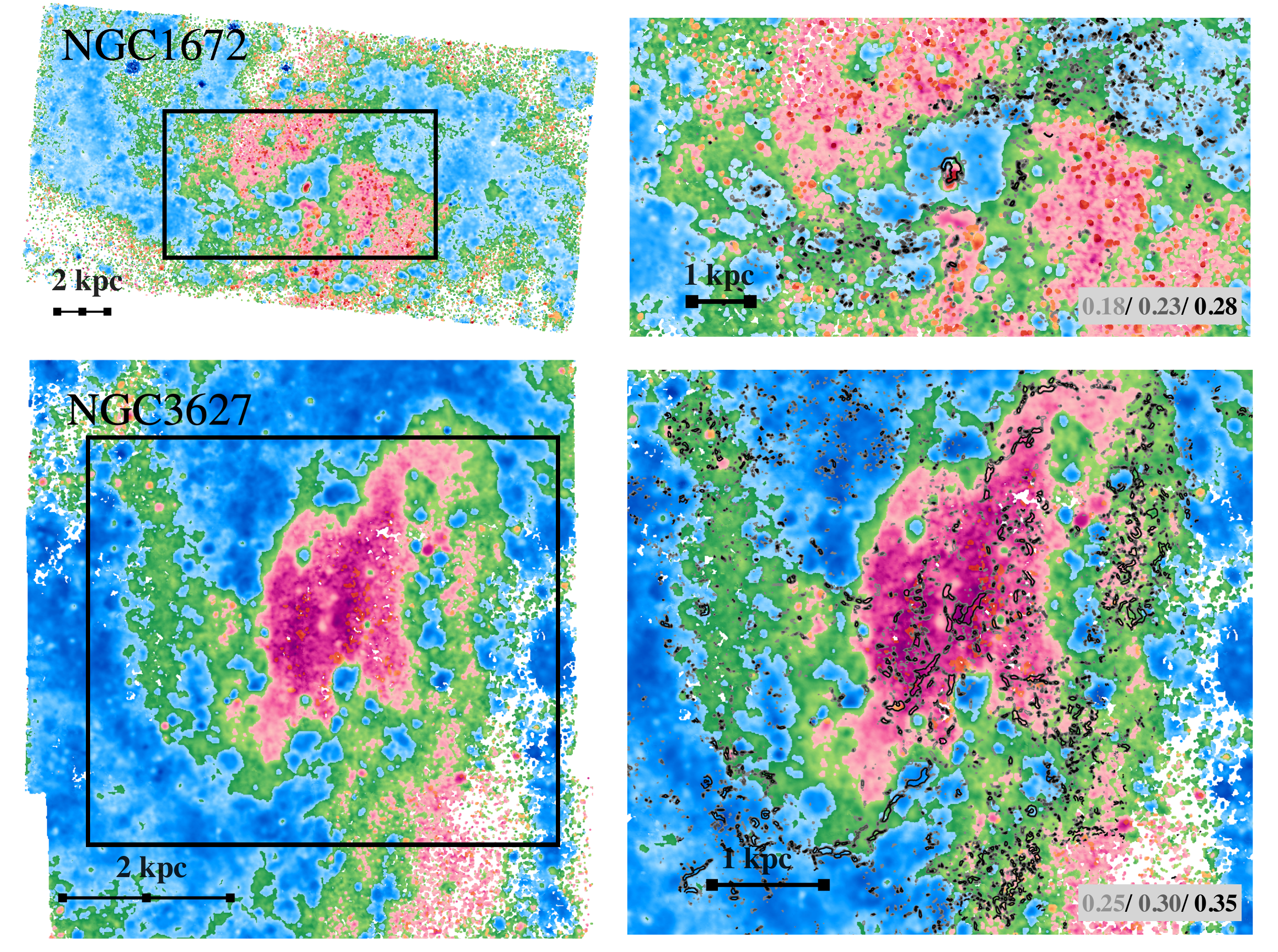} 
\caption{BPT maps of the sample galaxies. 
The right panels are zoomed-in versions of the BPT map overplotted with the velocity gradient contours. 
Numbers on the bottom right of each panel indicate the contour levels of velocity gradients in [\kmspc].
The color-coding follows the same as in Figure \ref{fig:bpt_diagram_1365}. Blue points indicate SF, Red to orange points denote Seyfert, magenta points indicate LINER, and green points represent composite region.}
\label{fig:vg_contour_on_bpt}
\end{figure*}

Dominant ionizing sources of gas can be separated using the BPT diagram (\citealt{baldwin_81, veilleux_87,kewley_01, kauffmann_03c}). 
A theoretical maximum starburst line (\citealt{kewley_01}, solid line in Figure \ref{fig:bpt_diagram_1365}) is defined by the upper limit of the theoretical pure stellar photo-ionization model. Points above this line are likely to be dominated by gas ionized by an AGN or shock.
Points below the dotted line in Figure \ref{fig:bpt_diagram_1365} (\citealt{kauffmann_03c}) represent regions where the emissions come from pure SF.
The regions between the dotted line and solid line indicate that gas is ionized by both SF and AGN, thus is named ``composite".
The dashed line (\citealt{cidfernandes_11}) divides Seyfert and low-ionization narrow emission-line regions (LINER).
We plot the BPT diagram for NGC 1365 in Figure \ref{fig:bpt_diagram_1365}. Each point denotes a spaxel of the galaxy. All the spaxels which satisfy the amplitude-over-noise ratio of emission-lines greater than 5 are plotted. Blue points represent emissions from star forming regions and green ones denote composite spectra. Orange to red points correspond to Seyfert and magenta points represent LINER. Points that are in close proximity to the maximum starburst line are plotted in lighter shades, whereas points distant from the maximum starburst line are denoted with darker shades.
The corresponding maps for all galaxies in the sample are presented in Figure \ref{fig:vg_contour_on_bpt} with the same color coding as in Figure \ref{fig:bpt_diagram_1365}.

NGC 1365 displays a biconical outflow (e.g., \citealt{phillips_83, jorsater_84, venturi_18, gao_21}). Within that outflow regions, emission-line ratios are located in either Seyfert or LINER area in the BPT diagram. Bar-driven high velocity gradient regions are mostly composite, LINER, or Seyfert.
The region marked by the yellow arrow in Figure \ref{fig:vg_contour_on_bpt} shows a remarkably sharp change compared to adjacent regions, from star-forming to composite and Seyfert type. However, in the other bar-driven high velocity gradient regions, the ionizing sources seem to be contaminated by the outflow.
Inside the bar radius of NGC 1512, the sources of ionization in most regions are LINER, and it also applies to strong shear regions driven by the bar. 
For NGC 1672, high shocks and shear regions are not preferentially occupied by any types of ionizing sources. 
In NGC 3627, the northern part of the high shear regions is mostly of composite or Seyfert type. However, the southern part of the high shear regions show either star forming or composite.

In the previous Section, we plot velocity gradients and SFR surface density in Figure 8. It contains all the spaxels of SF, LINER, composite, and AGN.  If we only consider SF spaxels (black contours in Figure 8), the mean SFR surface density is $\sim$5 times higher on average compared to the spaxels of composite, LINER, and Seyfert (purple contours in Figure 8). SF spaxels span slightly narrow range of velocity gradients with lower ($\sim$20$\%$ on average) mean value.

\subsection{SF classified with the BPT}
\begin{figure}[thb!]
\centering
\includegraphics[width=0.45\textwidth]{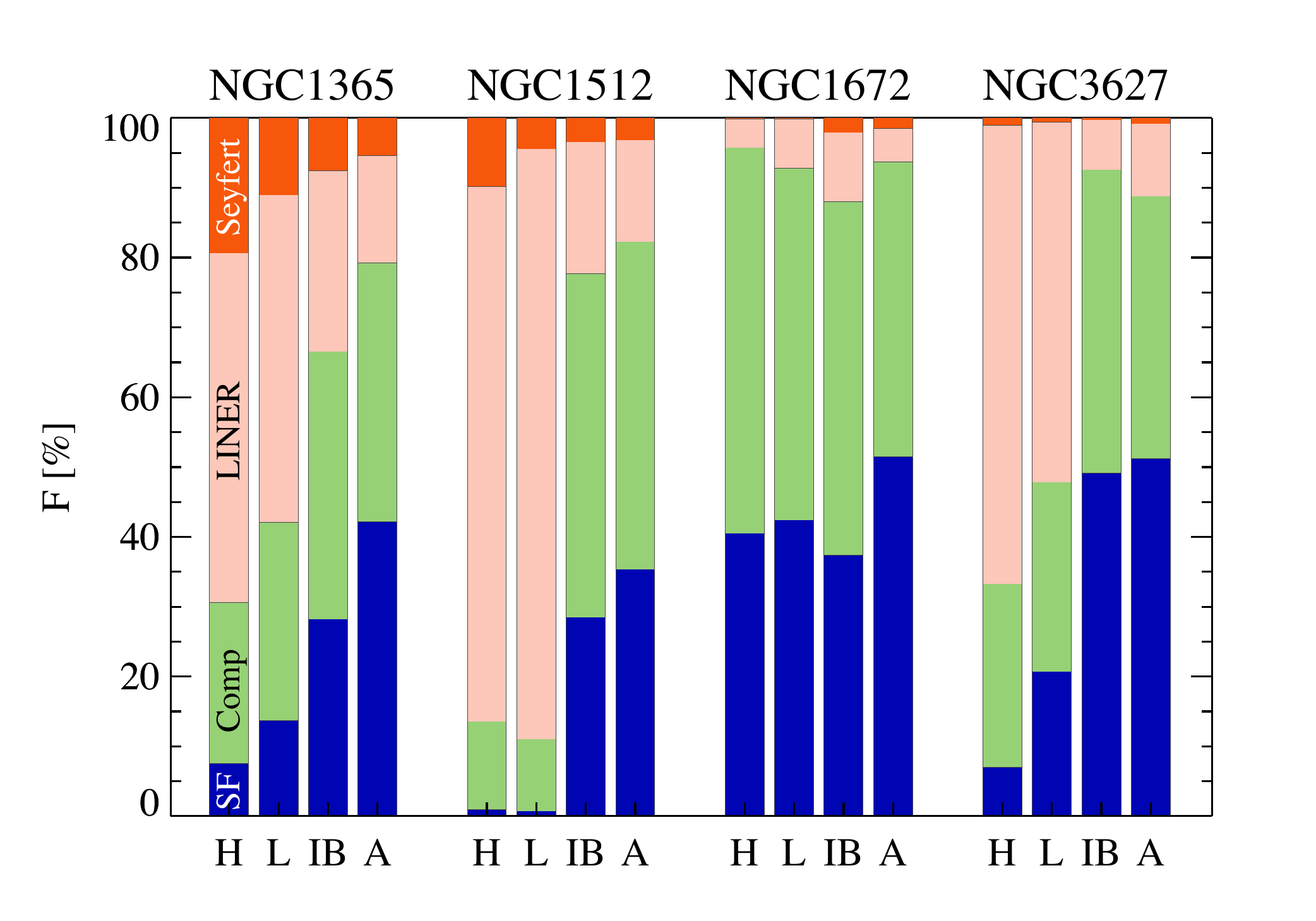} 
\caption{Ratio of ionizing sources in different structures of galaxies: high velocity gradient regions in the bar (H), low velocity gradient regions in the bar (L), interbar regions (IB) which is outside the bar but still within the bar radius, and all the other regions of the galaxy (A) for each object. 
Note that in this ``bar" region of ``H" and ``L", bar ends and nuclear regions are excluded to estimate the ratio to focus on regions of bar$-$driven shear and shock along the bar dust lane.}
\label{fig:ratio_bpt}
\end{figure}

In order to re-examine whether the SF is inhibited in high shear/shock regions, we classify the ionizing sources using the BPT diagnostics and compare with those of low shear/shock regions in the bar, and all the other regions of galaxy in Figure \ref{fig:ratio_bpt}.
Nuclear regions and bar ends are excluded in this ``bar'' region, as they exhibit active SF and shear and shocks due to bar-driven gas inflow are not strong there.
We define the high velocity gradient regions in the bar (H) as those with their velocity gradients greater than the upper 5th percentile in the bar for each galaxy, and define the low velocity gradient regions in the bar (L) to have their velocity gradients less than the upper 5th percentile. We also define the interbar region (IB), outside the bar but still within the bar radius.
We find that the fraction of the SF group in high velocity gradient regions in the bar (H in Figure \ref{fig:ratio_bpt}) is lower than that of low velocity gradient regions in the bar (L) and all the other regions of the galaxy (A). 
The majority of high velocity gradient regions exhibit composite or LINER. The fraction of SF group in high velocity gradient regions of the bar spans $\sim$ 2 -- 40$\%$.
This implies that SF is inhibited in high velocity gradient regions in the bar, which is in line with our results from the previous section that bar-driven shear/shock hinders SF.

By analyzing 240 barred galaxies with MaNGA, \citet{Krishnarao_20b} find that ionized gas is generally classified as LINER-like within the bars, composite in the interbar regions, while ionized gas from SF is located either at the center and/or outside the bar radius. Figure \ref{fig:ratio_bpt} also shows that in the interbar region (IB), the majority of spaxels are either composite or LINER which is consistent with the results of \citet{Krishnarao_20b}.
Except for NGC 1672, our sample galaxies fit well into their picture in general. However, some galaxies still show SF regions in the bar region (NGC 1365 and NGC 1672). We also found that there are Seyfert spaxels which are not just confined to the center, but related to the biconic outflow from the AGN (NGC 1365) or spread inside the bar radius (NGC 1512, NGC 1672) even without the AGN.

\subsection{Enhanced emission-line ratios in the high velocity gradient regions}
Our results imply that the emission-line ratios (e.g., [NII]/H$\alpha$ and [OIII]/H$\beta$) are elevated due to bar-driven shear/shock towards either composite or LINER in the BPT diagram.
However, for NGC 1512, the difference between high and low velocity gradient regions of the bar is negligible. There can be two reasons for this. 

First, the shock velocity and the fraction of shock converted to excite the gas play roles in enhancing the line ratio. \citet{ho_14} show that in their Figure 13, the line ratios of [NII]/H$\alpha$, [SII]/H$\alpha$ and [OI]/H$\alpha$ increase as the shock velocity increases. Also, within the same shock velocity, the line ratio of [OIII]/H$\beta$ increases as the shock fraction increases, where the shock fraction is the contribution of shock excitation relative to HII regions.
In Section~\ref{sec:velocityjump}, we show that the velocity jump in the high velocity gradient regions ranges from 75 \kms to 170 \kmsn, depending on the objects. Therefore, we would expect high shear regions to show enhanced [NII]/H$\alpha$ to some degree toward the LINER regime in the BPT diagram. But the shift of [NII]/H$\alpha$ also depend on the fraction of shock that is converted to change the emission-line ratio, which can not be readily estimated. Therefore, we infer that the shock fraction might not have been large enough to show a prominent change in the emission-line ratio for NGC 1512.

Second, even though the shear and shock may enhance the line ratio by shock-excitation, if the regions already have increased emission-line ratio and located in LINER or Seyfert domain, then it would be difficult to uplift the line ratios. The whole bar of NGC 1512 is dominated by the LINER or Seyfert type emissions judging by the BPT diagnostics.

It should be also noted that enhanced emission-line ratios do not necessarily guarantee that the ionizing source of the regions are shock originated. 
The leaking of ionizing photons from HII regions are suggested to be able to explain the diffuse ionized gas with LINER-like emissions (\citealt{belfiore_22}). Also, ionization by hot evolved stars (e.g., post AGB) are found to present LINER-like emission (\citealt{singh_13}). As post AGB stars are common in galaxies, their ionizing effect can be observed unless a strong radiation from SF or AGN outshines them. As we observe superimposed different ionizing sources along the line of sight, not just on the disk plane but also in the galactic halo, if there is lack of SF in the disk, we may well be left with the LINER-like emission from the post AGBs located along the line of sight. The whole bar regions of NGC 1512 may consist of old stars. 
Indeed, \citet{pessa_23} find that the mean mass weighted age of the stellar populations in the bar region of NGC 1512 is $\sim$ 10 Gyr.

Recently, \citet{johnston_23} proposed a novel method to distinguish ionizing sources by using the multi-dimensional diagnostic diagram which exploit the classical BPT, velocity dispersion, and equivalent width of H$\alpha$ (see also \citealt{cidfernandes_11, dagostino_19}). With the multidimensional diagnostics, they categorize more diverse range of ionization sources, such as the ionization from hot low-mass evolved stars, shocks, elevated kinematics with low emission-line ratio, diffuse ionized gas, in addition to the classical divisions of AGN, LINER, Seyfert, and SF.
Applying this multidimensional diagnostics to divide into more detailed classification is beyond the scope of the current paper as we focus on the impact of bar-driven shear and shock on star formation, and we will apply this in the upcoming future paper to explore further the impact of bar dynamics on gas ionization, gas properties such as virial parameter and turbulent pressure.

\section{Summary and Conclusion}  
\label{sec:summary}
We investigate the impact of bar-driven shocks and shear along the bar dust lane on SF and emission-line ratios by utilizing data from the PHANGS-MUSE and PHANGS-ALMA survey. Our results can be summarized as follows.

\begin{enumerate}
\item We derive velocity gradients using H$\alpha$ and stellar velocity maps, and find that velocity gradients are enhanced along the bar dust lane where shear and shock are expected to occur. CO intensity peaks coincide with the enhanced velocity gradient regions or slightly offset to the edge of the bar dust lane. Enhanced velocity gradient regions show velocity jumps from 50 to 150 \kmsn.

\item  We examine the velocity gradient versus SFR surface density and find that spaxels with high (low) molecular surface density show high (low) SFR surface density in general. In low velocity gradient regions, there are various SFR surface density values from low to high. However, as the velocity gradient increases, the range of SFR surface density narrows down such that the maximum SFR surface density decreases with velocity gradient. Exceptions are nuclear regions and star forming knots in the inner rings, which protrude out in the trend with enhanced velocity gradients.

\item  By comparing the H$\alpha$ emission-line maps and velocity gradient maps (Figure \ref{fig:VG_Ha_contour_set4}), we find that strong SF does not occur in the high velocity gradient regions.
We apply the BPT diagnostics on a spaxel by spaxel basis, and find that the fraction of SF spaxels is lower in high velocity gradient regions in the bar compared to low velocity gradient regions in the bar and to all the other regions of the galaxy (Figure \ref{fig:ratio_bpt}). This implies that SF is inhibited in the high velocity gradient regions in the bar where shocks and shear are strong.

\item The majority of high velocity gradient regions are classified into either LINER or composite categories. This may imply that emission-line ratios are elevated due to bar-driven shock or shear. But we cannot rule out that the absence of recent SF results in revealing LINER-like emission-line.

\item Our results are consistent with numerical simulations in which bars drive strong shear, and thus, inhibit SF (e.g., \citealt{emsellem_15}). Currently, it is not yet clearly known what properties of a barred galaxy control the strength of bar-driven shear and at what shear strength the SF is inhibited.
These questions might be answered through further numerical simulations which will lead us to understand the diverse impact of bars on the SF and galaxy evolution.
\end{enumerate}

\vspace{10mm}
This research was supported by the Basic Science Research Program through the National Research Foundation of Korea (NRF) funded by the Ministry of Education (RS-2023-00240212 and No. 2019R1I1A3A02062242) and NRF grant funded by the Korean government Ministry of Science and ICT (MSIT) (No. 2022R1A4A3031306). 
This work was supported by STFC grants ST/T000244/1 and ST/X001075/1. 
TK acknowledges the support from the Korea Foundation for Women In Science, Engineering and Technology (WISET) Grant (2022-804) funded by MSIT under the Support for Parental Leave Replacement Jobs.
MQ and PSB acknowledge support from the Spanish grants PID2019-106027GA-C44 and PID2019-107427GB-C31 respectively, funded by MCIN/AEI/10.13039/501100011033.
A.Z. acknowledges financial support by the research projects PID2020-113689GB-I00, and PID2020-114414GB-I00, financed by MCIN/AEI/10.13039/501100011033, the project A-FQM-510-UGR20 financed from FEDER/Junta de Andalucía-Consejería de Transforam-
ción Económica, Industria, Conocimiento y Universidades/Proyecto and by the grants P20{\_}00334 and FQM108, financed by the Junta de Andalucía (Spain).
JN acknowledges funding from the European Research Council (ERC) under the European Union’s Horizon 2020 research and innovation programme (grant agreement No. 694343). 
JMA acknowledges the support of the Viera y Clavijo Senior program funded by ACIISI and ULL and the support of the Agencia Estatal de Investigaci\'on del Ministerio de Ciencia e Innovaci\'on (MCIN/AEI/10.13039/501100011033) under grant nos. PID2021-128131NB-I00 and CNS2022-135482 and the European Regional Development Fund (ERDF) ‘A way of making Europe’ and the ‘NextGenerationEU/PRTR’.
AdLC  acknowledges financial support from the Spanish Ministry of Science and Innovation (MICINN) to the coBEARD project (PID2021-128131NB-I00). AdLC also acknowledges financial support from MICINN through the Spanish State Research Agency, under the Severo Ochoa Centres of Excellence Programme 2020-2023 (CEX2019-000920-S).
LPM thanks FAPESP through grant 2022/03703-1 and CNPQ through grant 307115/2021-6 for financial support.
LASL acknowledges CAPES for the support through grant 88887.637633/2021-00 and CNPQ through grant 200469/2022-3.
WTK was supported by the grant of the National Research Foundation of Korea (2022R1A2C1004810).
MGP acknowledges the support from the Basic Science Research Program through the National Research Foundation of Korea (NRF No-
2019R1I1A3A02062242 and NRF No-2018R1A6A1A06024970) funded by the Ministry of Education, Korea.
Based on observations collected at the European Southern Observatory under ESO programmes 1100.B-0651, 095.C-0473, and 094.C- 0623 (PHANGS-MUSE; PI: Schinnerer), as well as 094.B-0321 (MAGNUM; PI: Marconi), and 097.B-0640 (TIMER; PI: Gadotti). Raw and reduced data are available at the ESO Science Archive Facility. 

\vspace{5mm}
\facilities{VLT, ALMA, JWST}

\appendix
\section{Velocity gradient perpendicular to the bar and parallel to the bar}
\label{sec:app_vg}
\begin{figure}[ht]
\includegraphics[width=\textwidth]{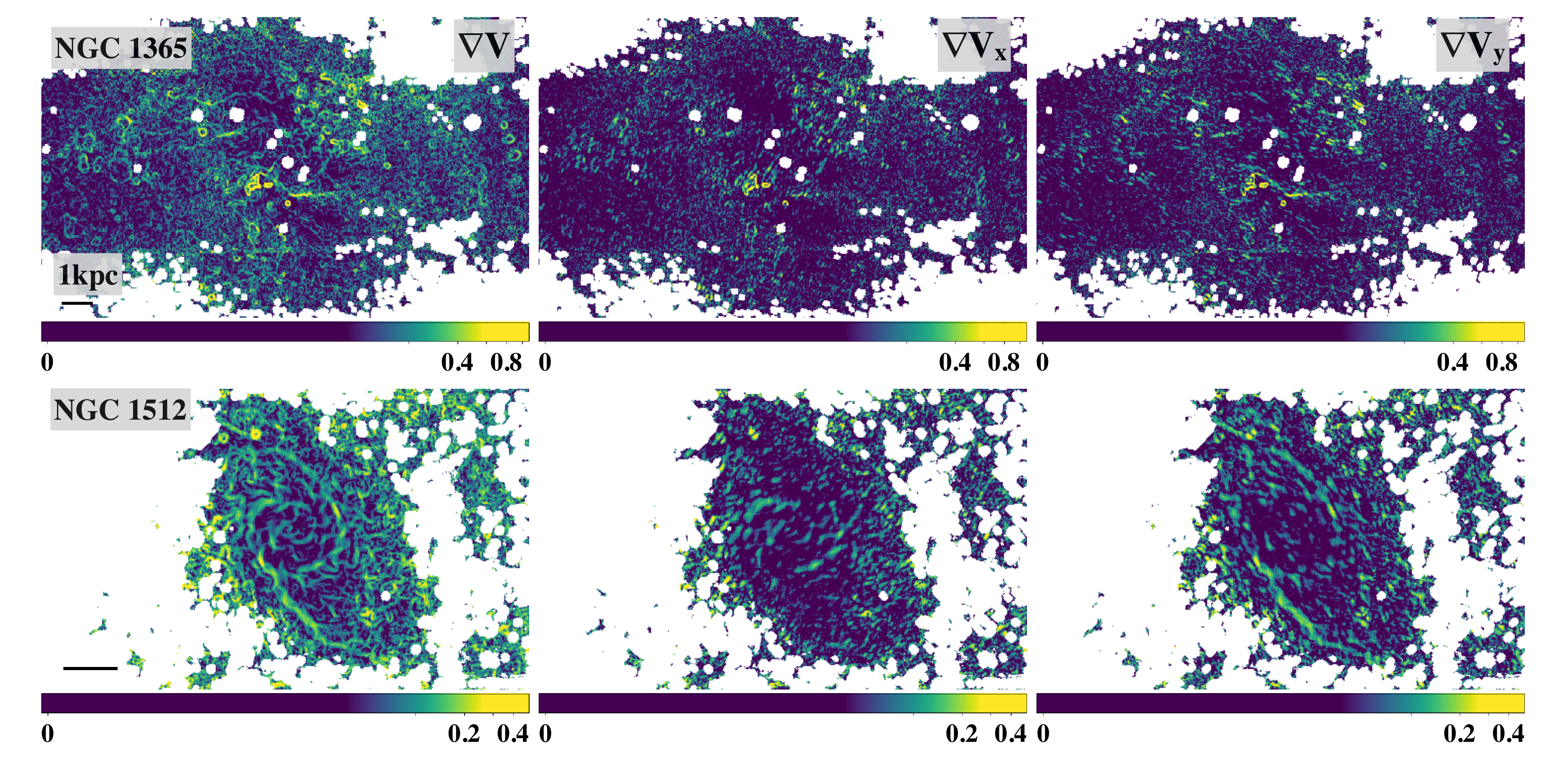}
\includegraphics[width=\textwidth]{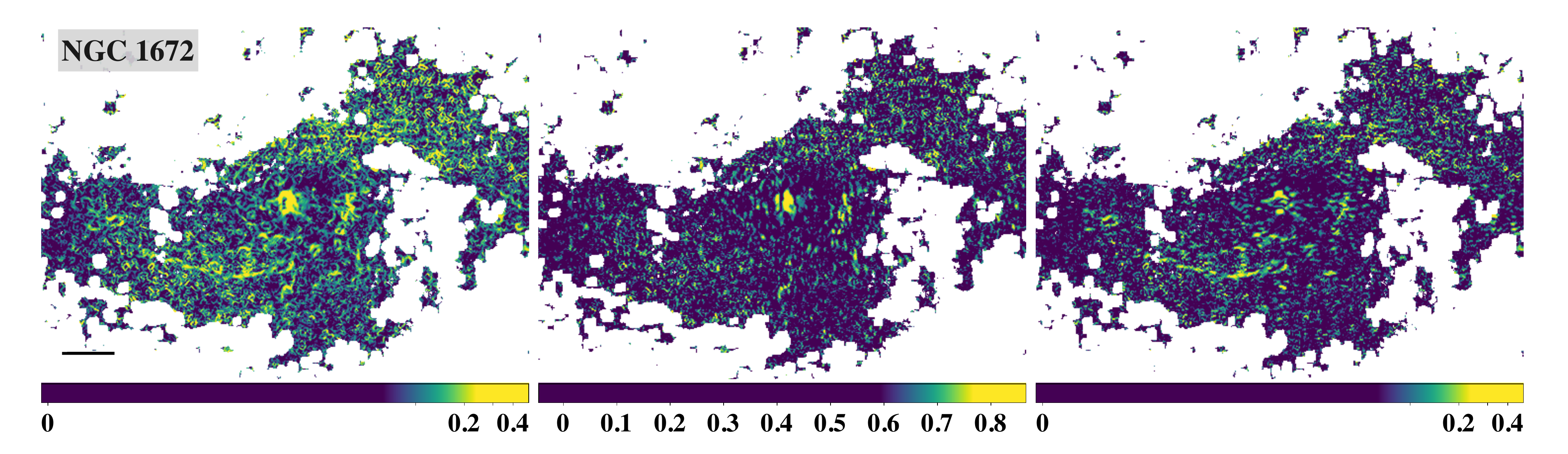}
\includegraphics[width=\textwidth]{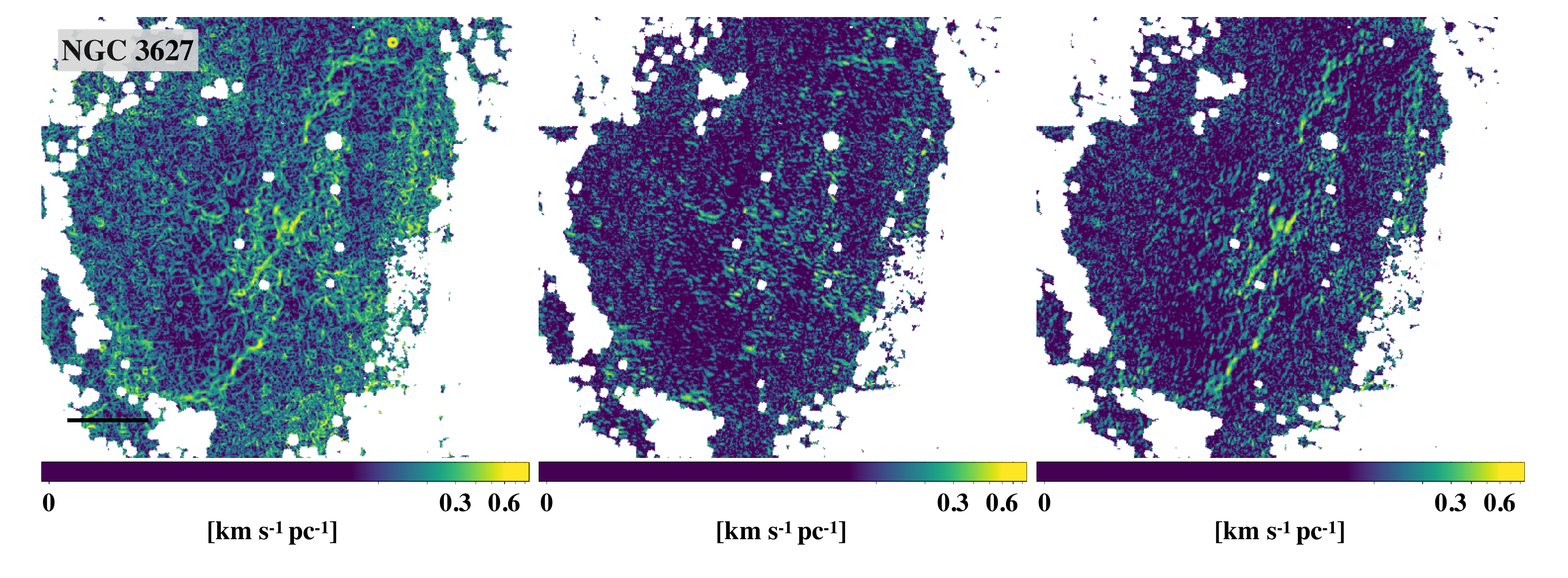}
\caption{The three columns show, respectively, the velocity gradients as in Section~\ref{sec:deriving_vg}, velocity gradients parallel component to the bar ($\nabla V_x$), and perpendicular to the bar ($\nabla V_y$) in unit of [\kmspc]. The black line on the bottom left of each panel spans 1 kpc.}
\label{fig:VG_per_par}
\end{figure}

Velocity gradients can be divided in two different components, perpendicular and parallel to the bar. We estimate parallel velocity gradient ($\nabla V_x$) and perpendicular velocity gradient ($\nabla V_y$) separately, and compare them in Figure \ref{fig:VG_per_par}.
We re-estimate the ratio of the SF group classified by the BPT diagnostics in high and low velocity gradient regions and all the other regions of each galaxy with $\nabla V_x$ and $\nabla V_y$ as in Figure \ref{fig:ratio_bpt}. We obtain the same results that the ratio of SF group is lower in high velocity gradient regions compared to the others.

\section{Comparing velocity gradients from H$\alpha$ and CO}
\label{sec:vg_co_ha}
\begin{figure}[ht]
\includegraphics[width=\textwidth]{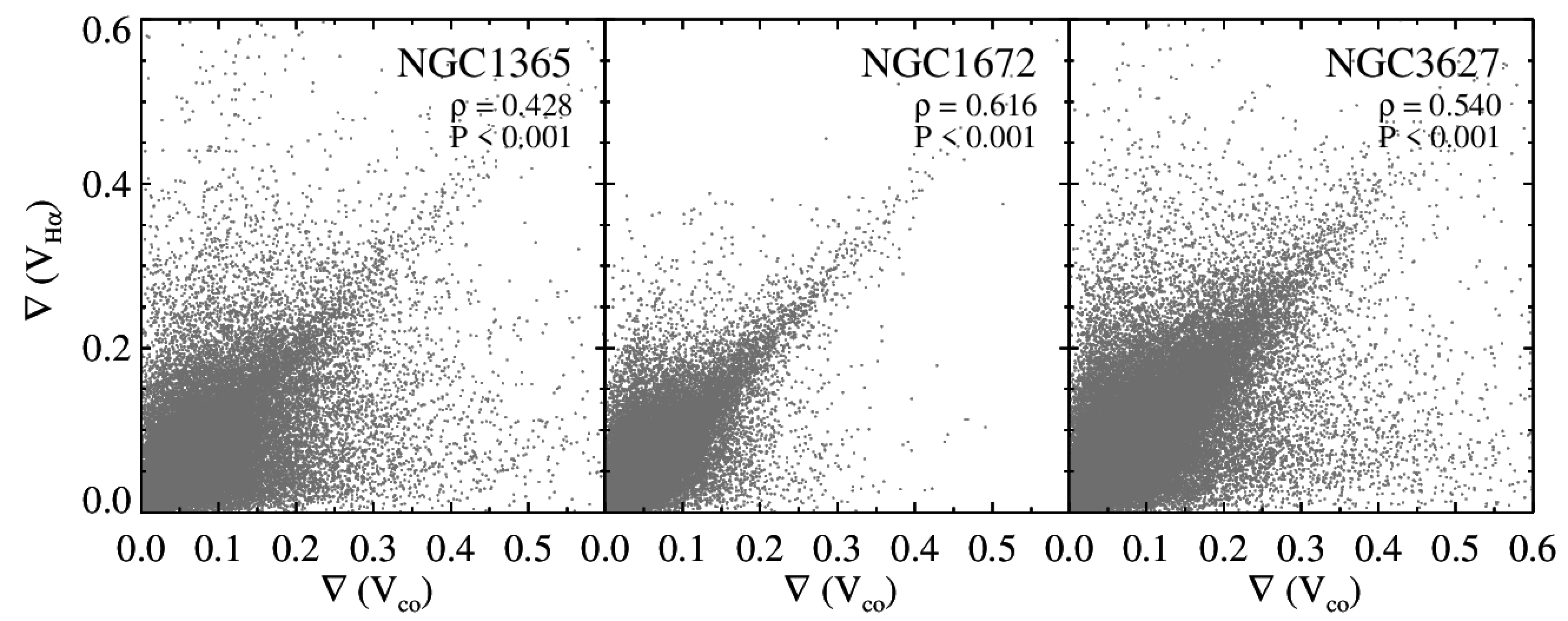}
\caption{Comparing velocity gradients from $V_{\rm H\alpha}$ and $V_{CO}$. Spearman coefficients are presented for each galaxy.}
\label{fig:VG_CO_Ha}
\end{figure}

Velocity gradients can also be obtained using the CO velocity field. However, compared to H$\rm \alpha$, CO is patchy and only detected marginally along the bar for some galaxies. 
In order to examine how well do H$\rm \alpha$ velocity gradients trace velocity gradients of molecular clouds, we compare velocity gradients from H$\rm \alpha$ ($\nabla V_{\rm H \rm \alpha}$) and CO ($\nabla V_{\rm CO}$) in Figure \ref{fig:VG_CO_Ha}. $\nabla V_{\rm H\alpha}$ is estimated from the map of ($V_{\rm H \alpha} - V_{\rm star}$) and $\nabla V_{\rm CO}$ is estimated with ($V_{\rm CO} - V_{\rm star}$) map following the same steps presented in Section~\ref{sec:deriving_vg}.
CO in the bar of NGC 1512 is marginally detected along the bar, therefore we could not obtain CO velocity gradients for this object. We find that while there is a spread (specially for low gradients), CO velocity gradients are correlated with H$\alpha$ velocity gradients. This suggests that we can use H$\rm \alpha$ velocity gradients to trace molecular gas velocity gradients for these galaxies.

\section{Velocity jumps for sample galaxies}
\label{sec:app_veljump}
\begin{figure*}[ht!]
\includegraphics[width=\textwidth]{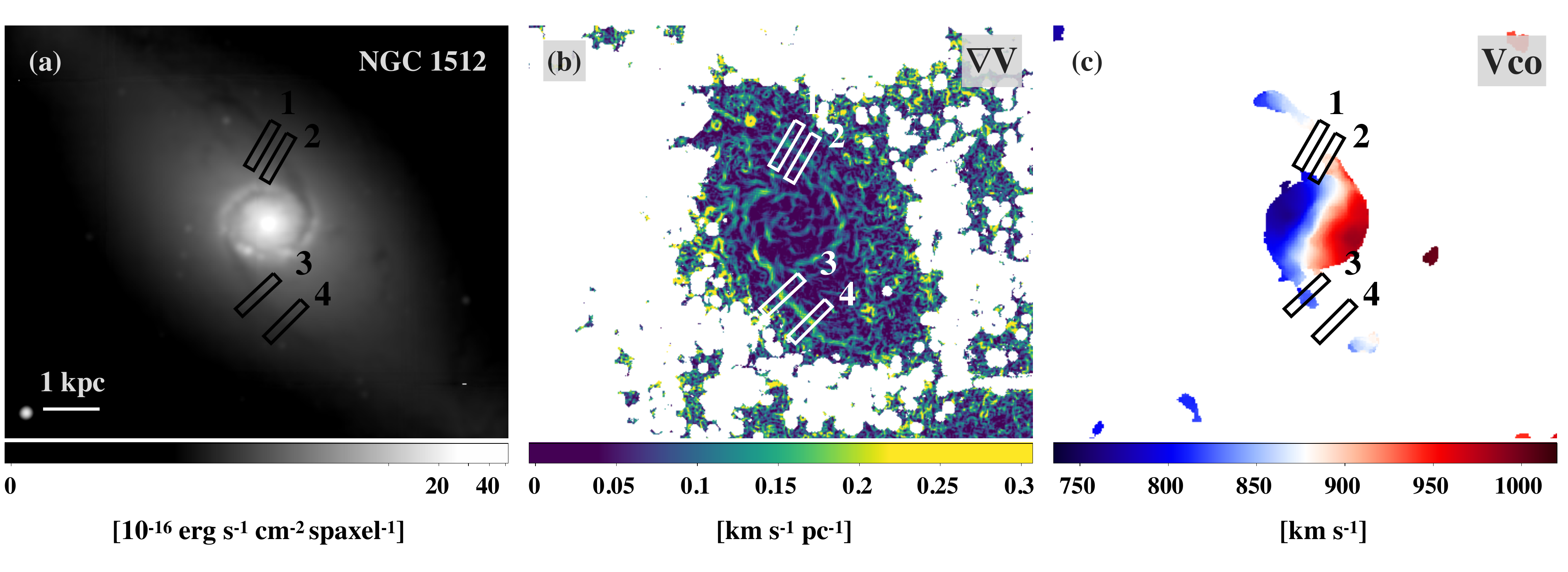} 
\includegraphics[width=\textwidth]{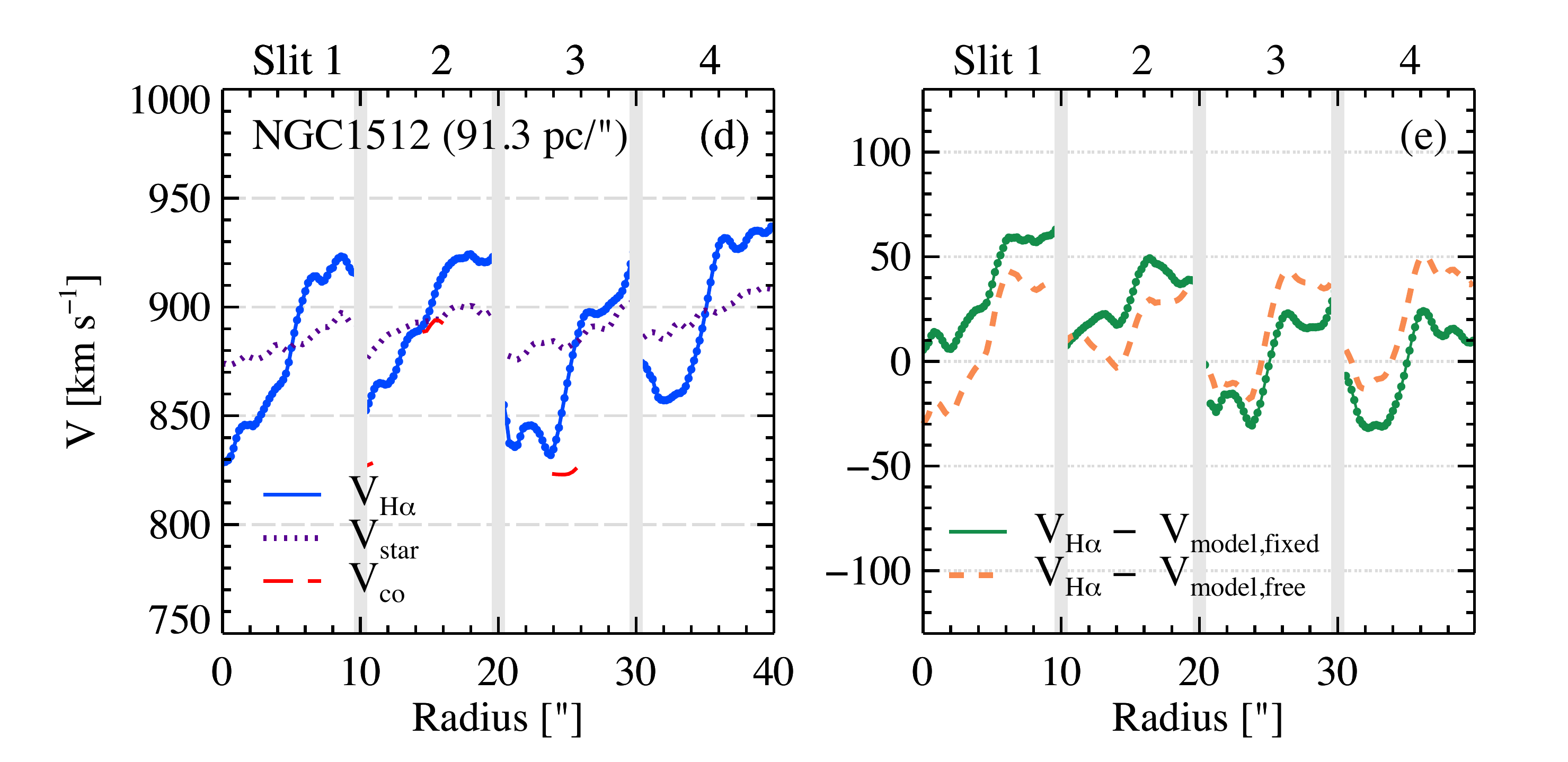} 
\caption{
Velocity jumps in high velocity gradient regions of NGC 1512. Four pseudo slits (0.6" $\times$ 10") are put perpendicular to the high velocity gradient segments as in the upper panels of the figure. (a): Optical image, (b): velocity gradients, (c): CO (2$-$1) velocity, (d) velocity jumps from south to north (upwards) along the slit of H$\alpha$ (blue solid line), CO (red dashed line, if available), and stellar velocity (purple dotted line), (e) velocity jumps after removal of the galaxy rotation model. Two cases of galaxy rotation models are obtained with {\sc 3D--BAROLO}, where position angles and inclinations are fixed (green solid line) and set to free (orange dashed line) for each ring. Velocity jumps reach 50 \kms after removal of the galaxy rotation (75 \kms after deprojection).}
\label{fig:vel_jump_appendix}
\end{figure*}

We present velocity jumps on high velocity gradient regions for the rest of the sample galaxies in Figure \ref{fig:vel_jump_appendix}.

\begin{figure*}[ht!]
\includegraphics[width=\textwidth]{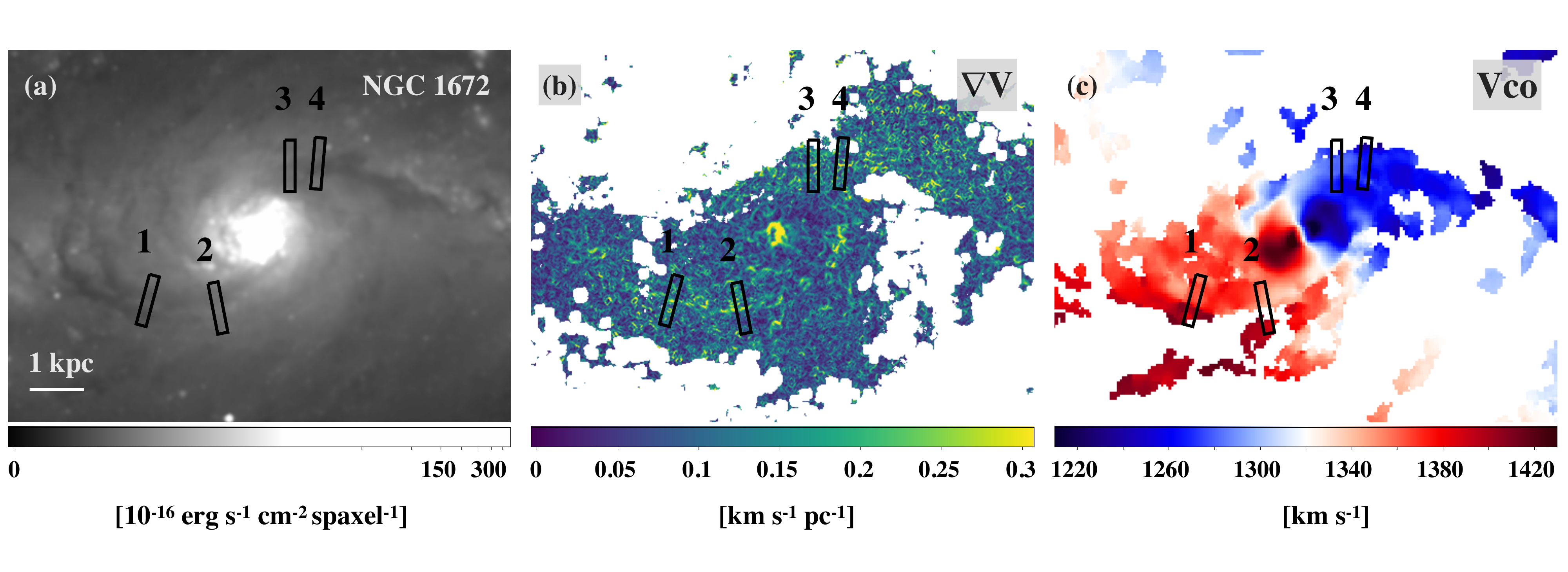} 
\includegraphics[width=\textwidth]{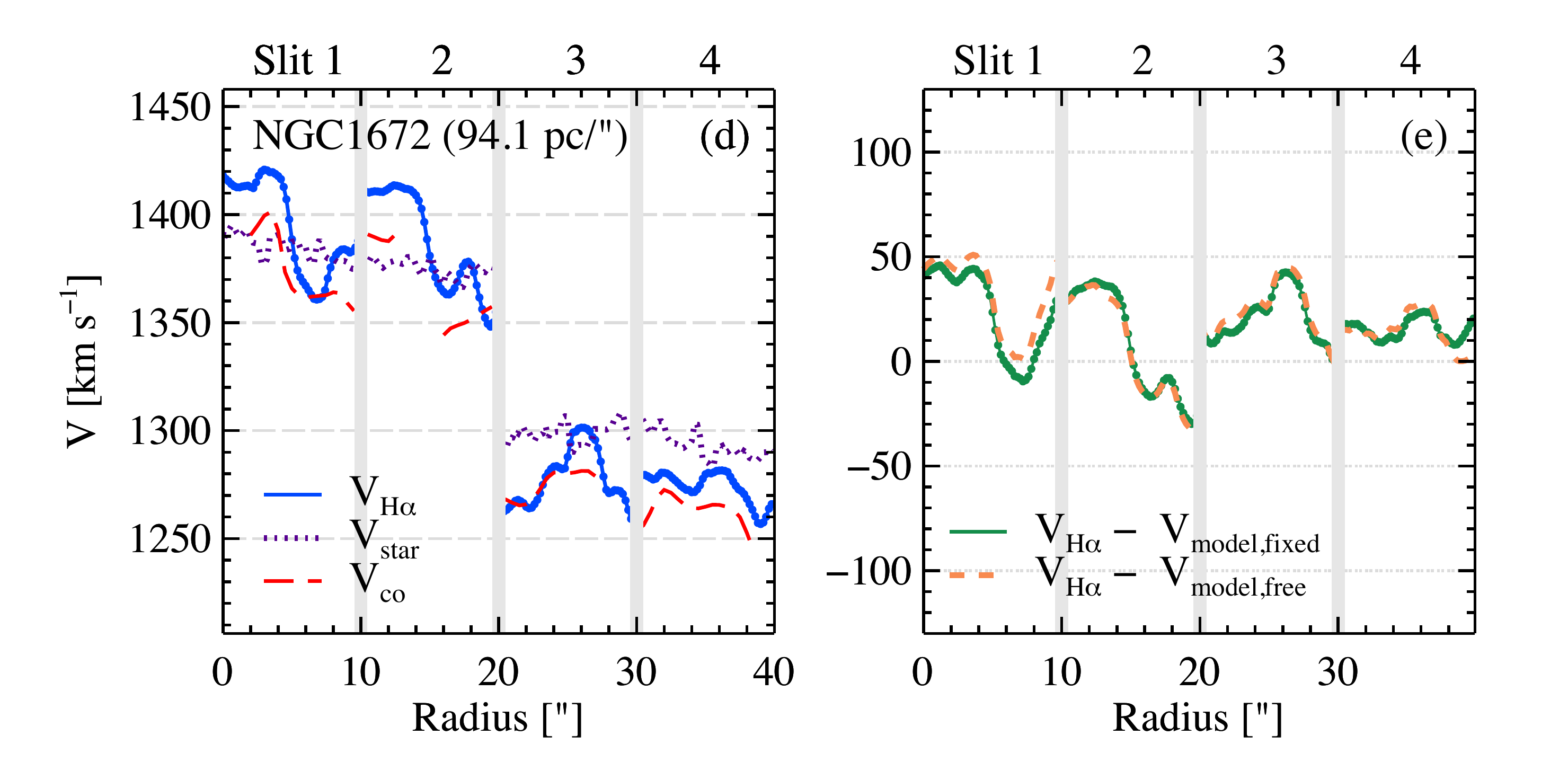} 
\caption{Velocity jumps for NGC 1672 (continued). 
Velocity jumps reach 50 \kms after removal of the galaxy rotation (75 \kms after deprojection).}
\end{figure*}

\begin{figure*}[ht!]
\includegraphics[width=\textwidth]{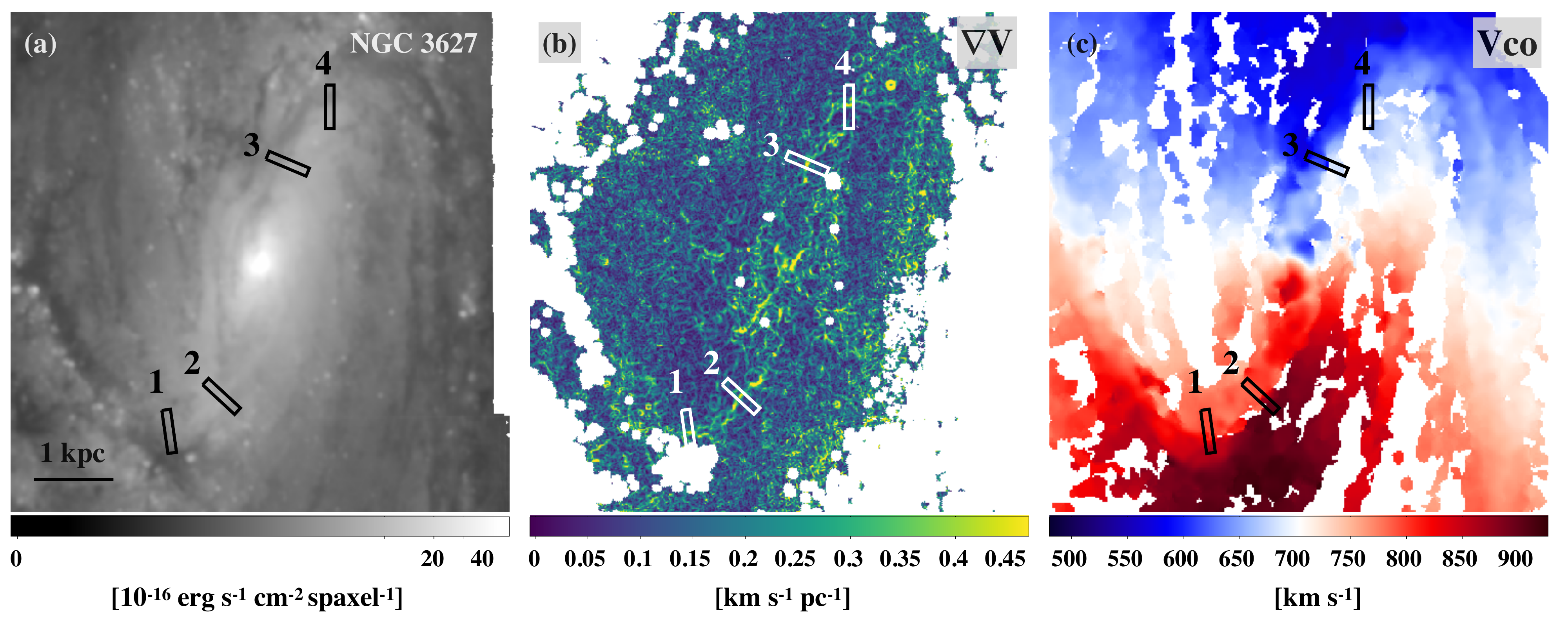} 
\includegraphics[width=\textwidth]{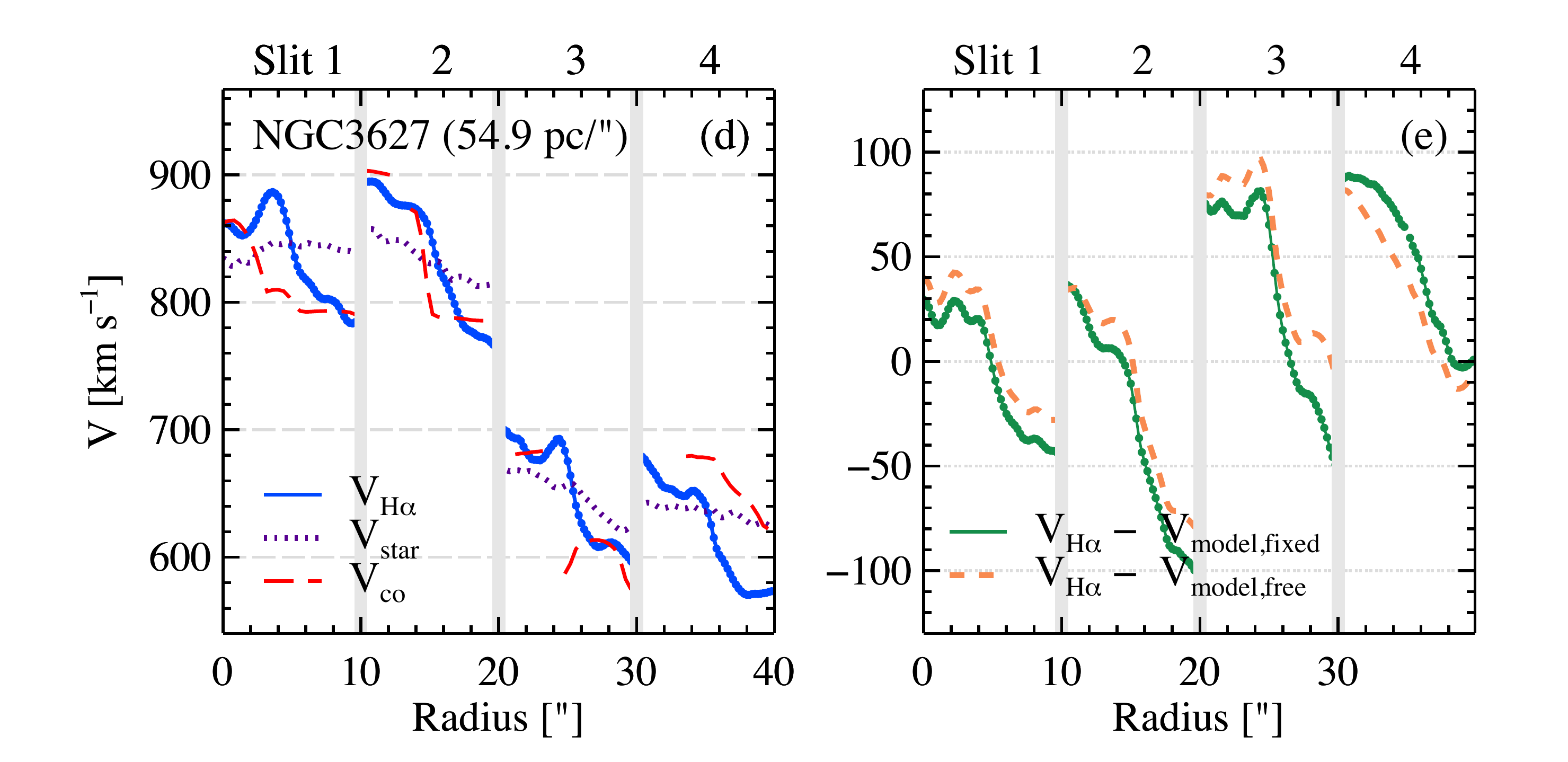} 
\caption{Velocity jumps for NGC 3627 (continued). 
Velocity jumps reach 80 \kms after removal of the galaxy rotation (95 \kms after deprojection).}
\end{figure*}


\section{Notes on individual galaxies} \label{sec:append_individuals}
\subsection{NGC 1365}
Although there are numerous enhanced velocity gradients which are thought to be originated by the expanding HII regions,
velocity gradients along the bar dust lanes are not prominent in the outer part (R$>$15'') of the galaxy, and are not symmetric. 
\citet{schinnerer_23} combined JWST/MIRI, ALMA and MUSE data and found that the gas inflow onto the central ring is asymmetric and different between the two dust lanes. In addition, they found that in the southern dust lane, young star clusters are rarely found even with abundant molecular gas with properties similar to the other dust lane are present. Exactly in this region (marked with yellow arrow in Figure \ref{fig:vg_contour_on_bpt}), we found that the velocity gradient is enhanced.


\subsection{NGC 1672}
In Figure \ref{fig:sfrsd_vg}, there are two branches that stick out prominently, one with high $\Sigma_{mol}$, and one with lower $\Sigma_{mol}$. The branch with high $\Sigma_{mol}$, where the $\Sigma_{mol}$ ranges between $10^{-1}$ and $10^0$ and the velocity gradients between 0.2 and 0.7 \kms, belong to the nuclear region (R$<$2.5'') inside the nuclear ring. In this region, CO (2$-$1) and H$\alpha$ intensity is high, and also the velocity gradient is extraordinarily high. 
The elevated velocity gradients may be caused by intense winds from young stars, as it appears to be the case in other galaxies with star-bursting nuclear rings, such as NGC 3351 (see \citealt{leaman_19}). The other branch with low $\Sigma_{\rm mol}$ ($10^{-3}\sim10^{-2}$) consist of regions around faint H$\alpha$ sources with distinctive velocities.

\subsection{NGC 3627}
Enhanced velocity gradients along the dust lane are disconnected at R$\sim$19" ($\sim$1 kpc) in both sides from the center (Figure \ref{fig:vg_set4}). The elevated velocity gradients in the southern outer part (R$>$19") are located at the leading side compared to the CO peaks (Figure \ref{fig:VG_CO_contour_set4}), but in the inner part (R$<$19''), they are located at the trailing side compared to the CO intensity. Along the northern part of the bar, enhanced velocity gradients are located along the trailing side. It is not clear why enhanced velocity gradient regions are located in the trailing side in some parts especially in this galaxy, as they are located along or at the leading side of the dust lane in other sample galaxies. Gas inflow is found to be reduced when there is a boxy/peanut-shaped bulge (\citealt{fragkoudi_16}), and interestingly the boxy/peanut-shaped bulge of the galaxy spans 18" (\citealt{erwin_16}). We suggest that the orbital structures due to the boxy/peanut-shaped bulge and spurs (\citealt{erwin_13, erwin_16}) may have an effect on the gas inflow, and may be related to the origin of this offset.

\bibliography{tkim_bar23}{}
\bibliographystyle{aasjournal}

\end{document}